\documentclass[aps,twocolumn,prl,preprintnumbers,showpacs,amsmath,amssymb,superscriptaddress]{revtex4-1}
\usepackage[dvipdfm, colorlinks, citecolor=blue]{hyperref} 
\usepackage{graphicx}
\usepackage{dcolumn}
\usepackage{threeparttable}
\usepackage{multirow}
\usepackage{booktabs}
\usepackage{txfonts}
\usepackage{xcolor}
\usepackage{bm}
\usepackage{amssymb}
\usepackage{amsmath}
\usepackage{latexsym}
\usepackage{epsfig}
\usepackage{amsbsy}
\usepackage{array}
\usepackage{tabularx}
\usepackage{esvect} 
\usepackage{extarrows}
\usepackage{float}
\usepackage[british]{babel}

\begin{document}

\title{Phase Transitions of Zirconia: Machine-Learned Force Fields Beyond Density Functional Theory}

\author{Peitao Liu}
\email{peitao.liu@univie.ac.at}
\affiliation{VASP Software GmbH, Sensengasse 8, 1090 Vienna, Austria}

\author{Carla Verdi}
\affiliation{University of Vienna, Faculty of Physics and Center for
Computational Materials Science, Kolingasse 14-16, A-1090, Vienna,
Austria}

\author{Ferenc Karsai}
\affiliation{VASP Software GmbH, Sensengasse 8, 1090 Vienna, Austria}

\author{Georg Kresse}
\affiliation{University of Vienna, Faculty of Physics and Center for
Computational Materials Science, Kolingasse 14-16, A-1090, Vienna,
Austria}
\affiliation{VASP Software GmbH, Sensengasse 8, 1090 Vienna, Austria}

\begin{abstract}
We present an approach to generate machine-learned force fields (MLFF) with beyond
density functional theory (DFT) accuracy.
Our approach combines on-the-fly active learning and $\Delta$-machine learning
in order to generate an MLFF for zirconia based on the random phase approximation (RPA).
Specifically, an MLFF trained on-the-fly during DFT based molecular dynamics simulations
is corrected by another MLFF that is trained on the differences between RPA and DFT calculated energies, forces and stress tensors.
Thanks to the relatively smooth nature of the differences, the expensive RPA calculations are performed only on a small number
of representative structures of small unit cells. These structures are determined by a singular value decomposition rank compression of the kernel matrix
with low spatial resolution.
This dramatically reduces the computational cost and allows us to generate an MLFF fully capable of
reproducing high-level quantum-mechanical calculations beyond DFT. We carefully validate our approach
and demonstrate its success in studying the phase transitions of zirconia.
\end{abstract}

\maketitle

Machine learning based regression techniques have become a prominent tool
to construct accurate interatomic potentials for materials modeling and
simulations~\cite{PhysRevLett.98.146401,PhysRevLett.104.136403,Botu_ML_review_JPCC2017,Behler_NN_review2017,PhysRevX.8.041048,
PhysRevB.90.024101,Shapeev_MTP2016,PhysRevB.97.184307,doi:10.1063/1.5020710, Michele_PCCP2016,Marques_npj2019, doi:10.1021/acs.jpclett.7b02010, PhysRevLett.114.096405,PhysRevLett.120.026102,PhysRevLett.122.225701}.
Machine-learned force fields (MLFF), however, are generally constructed by fitting the energies, forces, and
stress tensors derived by density functional theory (DFT) calculations, and therefore the accuracy of the resulting
MLFFs is largely limited by DFT. It is not surprising, then, that these MLFFs would fail in problems where DFT is inaccurate,
such as in systems where long-range electronic correlation effects play an important role.
This implies that pursuing an MLFF beyond DFT is highly desirable.

However, high-level quantum-mechanical (QM) methods such as the random phase approximation (RPA) are computationally much more
demanding than DFT, especially for structures containing many atoms. Hence, it is impractical to perform
these calculations for all structures in a typical MLFF training dataset
(including hundreds or thousands of supercell structures).
Several efforts have been made to circumvent this problem.
In Refs.~\cite{Anatole_2015JCTC,
Bartokek_SA2017, Chmiela2018,doi:10.1063/1.5078687, Smith2019,Bogojeski2020},
accurate but expensive high-level QM calculations
were performed on an affordable, reduced number of structures
in order to achieve near coupled cluster accuracy.
These studies, however, were mainly restricted to small molecules.
Although Refs.~\cite{MLTP_2019JCTC,Deringer2021} attempted to machine learn
the energies for condensed phase systems with near RPA accuracy,
no study so far has managed to train an MLFF that can predict
the forces as well as the stress tensors with the same level of
accuracy. This is indispensable for successful molecular dynamics (MD)
simulations of complex
phenomena at finite temperatures, such as solid-solid phase transitions.
In addition, the question remains on how to choose a small number of representative datasets
for the high-level QM calculations that ensures the desired accuracy.

In this Letter, we propose a general strategy to generate a kernel-based MLFF
capable of yielding RPA accuracy (not only in energies, but also in forces and stress tensors)
at a modest computational cost,
by combining an efficient on-the-fly active learning method~\cite{PhysRevLett.122.225701,PhysRevB.100.014105}
and a $\Delta$-machine learning ($\Delta$-ML) approach~\cite{Anatole_2015JCTC,Bartokek_SA2017}.
The success of $\Delta$-ML originates from the~\emph{ansatz} that low-level reference QM calculations
such as DFT already capture the most important contributions to the overall potential energy surface (though they might not be very accurate)
and therefore the remaining differences between high-level and low-level QM calculations become less corrugated
and thus easier to be machine-learned~\cite{Anatole_2015JCTC}.
This allows us to construct an accurate RPA-based MLFF for zirconia (ZrO$_2$),
with the computationally expensive RPA calculations performed only on a small number of representative structures of small unit cells,
significantly reducing the computational cost. We show that our
RPA-derived MLFF
accurately predicts structural parameters, phonon dispersions as well as the phase transition temperatures of zirconia.

\begin{figure}
\begin{center}
\includegraphics[width=0.48\textwidth, clip]{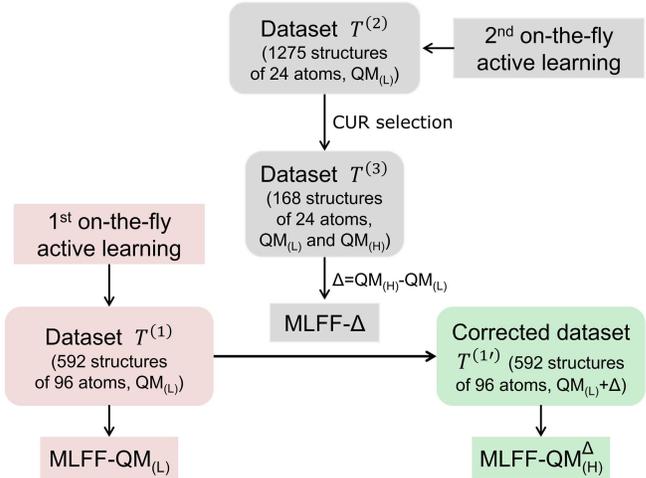}
\end{center}
\caption{Schematic workflow for the construction of high-level QM calculations based MLFF via the $\Delta$-ML approach.}
\label{fig:Work_flow}
\end{figure}

We start by describing the procedure that we propose for the construction of
an MLFF based on high-level QM
calculations via the $\Delta$-ML approach (MLFF-QM$_{\rm (H)}^\Delta$) (see Fig.~\ref{fig:Work_flow}).
($i$) First, an MLFF based on a low-level QM calculation (MLFF-QM$_{\rm (L)}$)
is trained on the fly during MD simulations.
Relatively large supercells of 96 atoms are used at this level.
We adopt Bayesian inference~\cite{PhysRevLett.122.225701,PhysRevB.100.014105}
to select 592 structures for the training dataset $T^{(1)}$, including
all three different phases (monoclinic, tetragonal and cubic) of ZrO$_2$.
Such a training dataset proved to be sufficient for generating an MLFF that can describe well the thermodynamic properties of ZrO$_2$~\cite{Carla_2021}.
For the detailed training strategy we refer to Ref.~\cite{Carla_2021}.
To highlight the power of the $\Delta$-ML approach, we purposely choose
the Perdew-Burke-Ernzerhof functional (PBE)~\cite{PhysRevLett.77.3865} as QM$_{\rm (L)}$,
since the PBE is found to be less accurate than other functionals for ZrO$_2$ by overestimating the lattice parameters
and energy differences between different phases (see Supplementary Material (SM) Table S1~\cite{SM}).
($ii$) Second, an auxiliary low-level MLFF is trained on-the-fly,
this time using smaller elongated supercells of 24 atoms.
The choice of such an elongated supercell ensures to include a certain amount of
long-range interactions (e.g., Van der Waals interactions) that shall be accounted for by the RPA.
The purpose of this step is to collect a second training dataset $T^{(2)}$ of small supercells
for which the RPA calculations are affordable and the generated MLFF in this step is discarded.
Eventually, 1275 structures are collected. The training details for this step are given in the SM~\cite{SM}.
($iii$) Third, a subset of structures (referred to as $T^{(3)}$) is selected from $T^{(2)}$
using the leverage-score CUR algorithm to perform a rank compression of the kernel matrix~\cite{Mahoney697,PhysRevB.100.014105}.
The resulting subset $T^{(3)}$ contains only 168 structures, when only pair descriptors
with low spatial resolution (0.8 $\AA$) and a small number of radial basis functions (8) are used to construct the kernel.
For these structures, low-level and high-level QM calculations are performed.
The differences in energies, forces and stress tensors between the high-level and low-level QM calculations
are then used to train a new MLFF (called MLFF-$\Delta$).
Following our \textit{ansatz}, this reduced set of training structures of small supercells should suffice to machine-learn the differences with a high accuracy.
($iv$) Finally, the energies, forces and stress tensors of the structures in $T^{(1)}$
are corrected by adding the differences predicted by the MLFF-$\Delta$. Using the updated $T^{(1)}$,
the MLFF-QM$_{\rm (H)}^\Delta$ is generated. This is supposed to be as accurate as
a force field that is directly machine-learned using high-level QM calculations.
We note that although in principle this final step can be omitted, the resulting two separate MLFFs
(i.e., MLFF-QM$_{\rm (L)}$ and MLFF-$\Delta$) will not have the same sort of convenience
as by combining both MLFFs into a single one.

\begin{table}
\caption {The validation root-mean-square errors (RMSE) in energies per atom (meV/atom), forces (eV/\AA) and stress tensors (kbar)
for different MLFFs. The test dataset includes 120 structures of 96 atoms (see SM~\cite{SM}). Here, $\Delta$=SCAN$-$PBE.
}
\begin{ruledtabular}
\begin{tabular}{lccc}
 & Energy & Force & Stress  \\
 \hline
MLFF-PBE                & 2.40 & 0.135 & 2.29 \\
MLFF-SCAN               & 2.49 & 0.139 & 2.38 \\
MLFF-SCAN$^\Delta$      & 2.37 & 0.139 & 2.30 \\
MLFF-$\Delta$           & 0.30 & 0.010 & 0.24 \\
\end{tabular}
\end{ruledtabular}
\label{tab:error_MLFF-DFT}
\end{table}

In the following, we validate our proposed scheme by first taking the
strongly constrained appropriately normed (SCAN)~\cite{PhysRevLett.115.036402} and PBE functionals as an example,
where SCAN and PBE are regarded as high-level and low-level QM methods, respectively.
As shown in Table~\ref{tab:error_MLFF-DFT}, all generated MLFFs are very accurate with small training and validation errors.
In particular, MLFF-SCAN$^\Delta$ derived by the $\Delta$-ML approach exhibits almost the
same accuracy as MLFF-SCAN, which was directly trained by SCAN.
In addition, we find that MLFF-SCAN$^\Delta$ performs almost equally well as MLFF-SCAN in predicting structural and vibrational properties for each phase of ZrO$_2$ (see SM Table S1 and Fig.~S4~\cite{SM}), validating the feasibility of the $\Delta$-ML approach.
It should be stressed that for training MLFF-SCAN$^\Delta$,
the SCAN calculations were performed solely on the $T^{(3)}$ dataset including only 168 structures of 24 atoms.
This significantly reduces the computational cost as compared to MLFF-SCAN,
which was directly trained on the $T^{(1)}$ dataset including 592 structures of 96 atoms.
We note that it is possible to further reduce the number of structures in $T^{(3)}$ for machine-learning the differences,
without significantly reducing the accuracy of the resulting MLFFs (see SM Table S2~\cite{SM}).

\begin{table}
\caption {The validation RMSE in energies per atom (meV/atom), forces (eV/\AA) and stress tensors (kbar)
calculated by MLFF-RPA$^\Delta$ on a test dataset including 60 structures of 24 atoms (see SM~\cite{SM}).
For comparison, the errors for MLFF-PBE and MLFF-SCAN are also given.
}
\begin{ruledtabular}
\begin{tabular}{lccc}
 & Energy & Force & Stress  \\
 \hline
MLFF-RPA$^\Delta$      &   3.77  &   0.136  &   5.47 \\
MLFF-PBE               &   3.68  &   0.129  &   4.71 \\
MLFF-SCAN              &   3.70  &   0.132  &   4.89 \\
\end{tabular}
\end{ruledtabular}
\label{tab:error_MLFF-RPA}
\end{table}

\begin{figure*}
\begin{center}
\includegraphics[width=0.85\textwidth, clip]{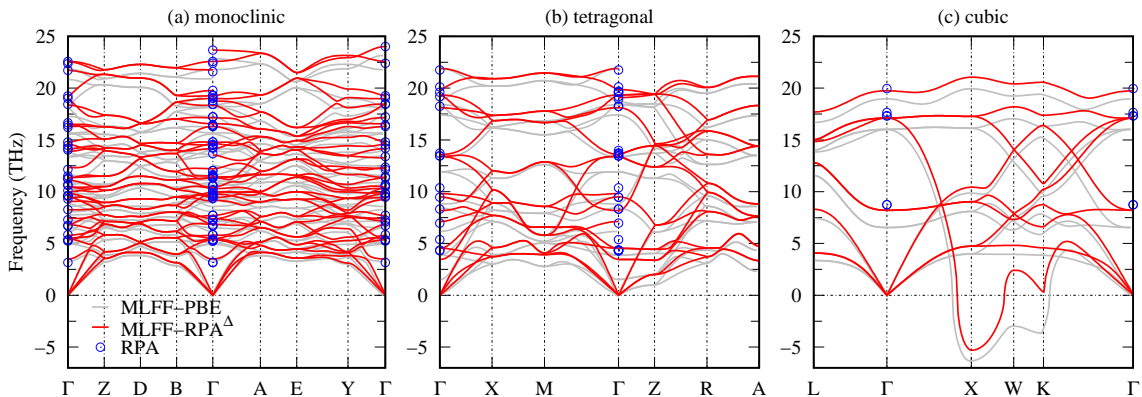}
\end{center}
\caption{Phonon dispersion relations of (a) monoclinic, (b) tetragonal, and (c) cubic  ZrO$_2$  at 0 K predicted by MLFF-RPA$^\Delta$ (red lines).
Direct RPA calculated phonon frequencies at $\Gamma$ are shown as blue circles.
The results from MLFF-PBE are also displayed (grey lines).}
\label{fig:phonon_MLRPA}
\end{figure*}

We now extend the application of the $\Delta$-ML approach to the construction of an RPA-derived MLFF.
For the structures in the dataset $T^{(3)}$, the RPA energies and forces are calculated using an efficient low-scaling algorithm~\cite{PhysRevB.90.054115,PhysRevLett.118.106403}.
The stress tensors at the RPA level are obtained via finite differences (see SM~\cite{SM} for details).
Due to the large computational cost of the RPA calculations, the resulting MLFF-RPA$^\Delta$ is validated on a reduced test dataset consisting of 60 structures of 24 atoms.
The validation errors are shown in Table~\ref{tab:error_MLFF-RPA}.
MLFF-RPA$^\Delta$ exhibits comparable errors as MLFF-PBE and MLFF-SCAN, implying comparably good accuracies.
One may also notice that, as compared to the 96-atom cells,  all the MLFFs exhibit relatively larger RMSEs for the energy per atom
and stress tensors on smaller unit cells of 24 atoms, whereas the RMSEs for forces
remain almost unchanged (compare Tables~\ref{tab:error_MLFF-RPA} and \ref{tab:error_MLFF-DFT}).
This can be understood from the error propagation with respect to the system size.
Specifically, assuming that the errors in the predicted local energies are statistically independent,
the RMSEs of the energy {\em per atom} and stress tensors will decrease by a factor
$1/\sqrt{N}$ if the system becomes $N$ times larger.
However, for the RMSE of forces, this error propagation rule does not apply, since the force is an intensive property that is independent of system size.
For a more detailed discussion on the error propagation with respect to the system size, we refer to the SM~\cite{SM}.

After validating the MLFF-RPA$^\Delta$  on a test dataset, we turn to its prediction of
ground-state properties such as lattice parameters, the energy differences between the three phases,
and the phonon dispersion relations of ZrO$_2$.
We find that MLFF-RPA$^\Delta$ yields an excellent description of the lattice parameters of the three phases, on par with or even slightly better than SCAN (see SM Table S1~\cite{SM}).
This is expected, since both SCAN and RPA account for certain medium-range electron correlations
and SCAN has been shown to be very close to RPA in the prediction of lattice parameters~\cite{PhysRevMaterials.3.103801}.
Similarly, we observe only small differences between MLFF-RPA$^\Delta$ and SCAN in the
predicted phonon dispersions (see SM Fig.~S5~\cite{SM}).
However, MLFF-RPA$^\Delta$  predicts smaller energy differences between the phases than SCAN.
Our results are consistent with Ref.~\cite{PhysRevResearch.2.043361},
which shows that many-electron calculations such as RPA or coupled cluster singles and doubles theory
yield smaller energy differences than DFT for ZrO$_2$.
To further validate the accuracy of MLFF-RPA$^\Delta$, we show that the energy differences between the three phases, as well as
the phonon frequencies at $\Gamma$, calculated directly using the RPA are in very good agreement with the predictions
by MLFF-RPA$^\Delta$ (see Table~\ref{tab:lattice_constants_Main} and Fig.~\ref{fig:phonon_MLRPA}).

\begin{figure*}
\begin{center}
\includegraphics[width=1.0\textwidth, clip]{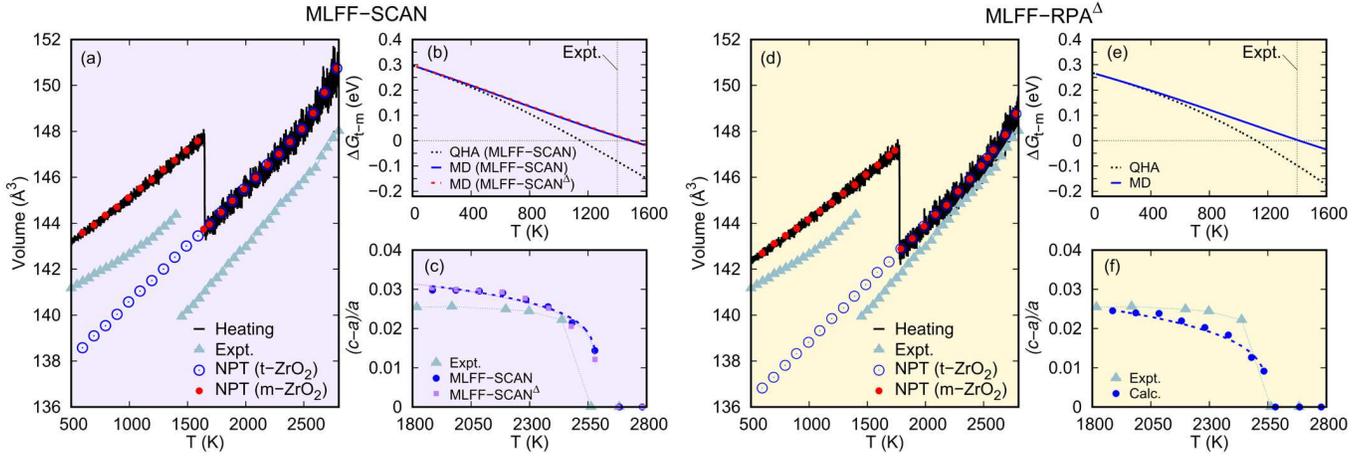}
\end{center}
\caption{
Phase transitions of ZrO$_2$ from MD simulations and thermodynamic integration.
 (a) and (d): Evolution of the unit-cell volume with temperature during a heating
 simulation (heating rate 0.5 K/ps) of a 324 atom supercell superimposed by MD simulations at fixed temperatures starting from m-ZrO$_2$ or t-ZrO$_2$ phases.
 (b) and (e): Free energy difference (per unit cell) between the tetragonal
 and monoclinic phase $\Delta G_\textup{t-m}$  as a function of temperature predicted by the QHA using classical
 Maxwell-Boltzmann statistics as well as by fully anharmonic MD calculations.
 (c) and (f) The tetragonal distortion $(c-a)/a$ and its fit to the function $(T_\textup{c}-T)^\alpha$.
(a), (b), and (c) are obtained using MLFF-SCAN, while (d), (e), and (f)  are obtained using MLFF-RPA$^\Delta$.
The values of $\Delta G_\textup{t-m}$ and $(c-a)/a$ predicted by MLFF-SCAN$^\Delta$  are also shown in (b) and (c), respectively.
The experimental data are taken from Refs.~\cite{kisi1998,https://doi.org/10.1111/j.1151-2916.1985.tb15247.x}.
}
\label{fig:RPA_vs_SCAN}
\end{figure*}

\begin{table}
  \caption{ Zero-temperature volumes ($\AA^3$/f.u.) and energy differences (eV/f.u.) between
  phases predicted by SCAN, MLFF-SCAN and MLFF-RPA$^{\Delta}$.
 The values in parentheses are RPA predicted energy differences for the structures that are optimized by MLFF-RPA$^{\Delta}$.
The predicted transition temperatures ($T_c$ in K) and transition enthalpies for the tetragonal and monoclinic phases ($\Delta H_{t-m}$ in eV/f.u.)
are given and compared to the experimental values~\cite{MORIYA2006211,kisi1998,https://doi.org/10.1111/j.1151-2916.1985.tb15247.x}.
The experimental volumes are extrapolated to 0 K~\cite{PhysRevB.49.11560,https://doi.org/10.1111/j.1151-2916.1985.tb15247.x}.
Full information including the structural parameters is given in SM Table S1~\cite{SM}.
}
  \begin{ruledtabular}
 \begin{tabular}{lcccccc}
                           & SCAN   & MLFF-SCAN & MLFF-RPA$^{\Delta}$& Expt. \\
                           \hline
Monoclinic                  &        &       &         &       \\
Volume                      &  35.35 & 35.37 &  35.20  & 35.22 \\
\\
Tetragonal                  &        &       &         &       \\
Volume                      &  33.82 & 33.90 &  33.47  & 33.01 \\
$\Delta E_{t-m}$            &  0.074 & 0.074 &  0.067 (0.069)  & --- \\
$\Delta H_{t-m}$            &  ---   & 0.069 &  0.069  & 0.056$\pm$0.003\\
$T_c (t-m)$            &  ---   & 1492 &  1415  & 1400 \\
\\
Cubic                       &        &       &         &       \\
Volume                      &  32.92 & 32.97 &  32.70  & --- \\
$\Delta E_{c-t}$            &  0.085 & 0.083 &  0.053 (0.047)  & --- \\
$T_c (c-t)$            &  ---   & 2585 &   2546  & 2570
 \end{tabular}
 \end{ruledtabular}
 \label{tab:lattice_constants_Main}
\end{table}

With our accurate MLFF-RPA$^\Delta$ in hand, we are now in a position to study the phase transitions of ZrO$_2$.
At ambient pressure, pure ZrO$_2$ exposes three structural phases. At high temperature it adopts a cubic structure,
which transforms to the tetragonal structure at about 2570 K~\cite{https://doi.org/10.1111/j.1151-2916.1985.tb15247.x}.
Around 1400 K the structure then undergoes a tetragonal to monoclinic phase transition~\cite{kisi1998}.
Let us first start by calculating the phase transition temperature ($T_c$) from the monoclinic to tetragonal phase
using the quasi-harmonic approximation (QHA).
MLFF-PBE predicts a value of 1511 K for $T_c$, 111 K larger than the experimental value.
MLFF-SCAN$^\Delta$ and MLFF-SCAN yield very close values of $T_c$, about 1148 K and 1164 K, respectively.
MLFF-RPA$^\Delta$  predicts a slight lower value of 1117 K.
In general, we find that within the QHA the predicted $T_c$ is correlated to the calculated energy differences
between the two phases at 0 K (see SM~\cite{SM}).

The QHA only partially takes into account anharmonic effects via the volume dependence of the vibrational frequencies.
To fully account for the anharmonicity, we performed MD simulations using the MLFFs.
We restrict our discussions only to MLFF-SCAN and MLFF-RPA$^\Delta$.
The evolution of the system volume with temperature predicted by MLFF-SCAN and MLFF-RPA$^\Delta$
are illustrated in Figs.~\ref{fig:RPA_vs_SCAN}(a) and (d), respectively.
As in the experiment~\cite{kisi1998}, the first-order transition between the monoclinic and tetragonal phase,
manifested by a sharp change in the volume, is observed in both simulations.
However, there is no obvious volume discontinuity in the tetragonal to cubic transformation,
but only a small change in the slope of the thermal expansion, indicating a second-order nature of the phase transition.
We also notice that the volumes predicted by MLFF-RPA$^\Delta$ are in better agreement with
experiment than MLFF-SCAN in the entire temperature range, and especially so for the high-temperature structures.
However, the $T_c$ predicted by direct MD heating simulations for both MLFFs are overestimated by about 250 K compared to experiment.
Moreover, we find that upon cooling the tetragonal to monoclinic phase transition is not reversible.
This makes it impossible to mitigate the error in estimating $T_c$ by averaging the transition temperatures
obtained from heating and cooling runs~\cite{Peitao_2020}.

To accurately determine the theoretical $T_c$, we followed the thermodynamic integration method developed in Ref.~\cite{Carla_2021}.
Specifically, the fully anharmonic free energy ($G$) of the monoclinic and tetragonal phase as a function of temperature is calculated as~\cite{Carla_2021}
\begin{equation} \label{eq:free-ene}
 G(T)=-T\int_{T_0}^T \frac{H(T')}{T'^2}\,\mbox{d}T'+\frac{G_0}{T_0}T,
\end{equation}
where $H=U+PV$  is the enthalpy with $U$ being the internal energy of the system, and $G_0$  is the Gibbs free energy at temperature $T_0$.
We performed the integral from $T_0$=25 K up to 1600 K with $G_0$ obtained from the QHA.
The integration paths are continuous, because the
tetragonal phase is metastable and does not transform into the monoclinic phase during our
MD simulations, while the monoclinic phase remains stable up to 1600 K [see Figs.~\ref{fig:RPA_vs_SCAN}(a) and (d)].

The free energy difference between the tetragonal and monoclinic phase
as a function of temperature is shown in  Figs.~\ref{fig:RPA_vs_SCAN}(b) and (e)
for MLFF-SCAN and MLFF-RPA$^\Delta$, respectively. The results obtained from Eq.~\eqref{eq:free-ene}
are compared to the ones calculated within the QHA using classical Maxwell-Boltzmann statistics.
According to these free energy calculations, $T_c$ is about 1492 K and 1415 K for MLFF-SCAN
and MLFF-RPA$^\Delta$, respectively, in excellent agreement with the experimental value.
As expected, MLFF-SCAN$^\Delta$ predicts almost an identical $T_c$ as MLFF-SCAN [see Fig.~\ref{fig:RPA_vs_SCAN}(b)].
In comparison, the QHA underestimates the $T_c$ by about 252 K and 283 K for MLFF-SCAN and
MLFF-RPA$^\Delta$, respectively, highlighting the need to account for the anharmonicity beyond the QHA.

The tetragonal to cubic phase transition can be more straightforwardly described using direct MD simulations.
In experiments the nature of this transition is not unambiguous, because cubic ZrO$_2$ is observed
only at very high temperatures above 2570 K~\cite{https://doi.org/10.1111/j.1151-2916.1985.tb15247.x}.
This makes experimental studies difficult.
From our MD simulations, we observe a continuous transition without thermal hysteresis.
In addition, we observe frequent fluctuations between the two phases near the transition temperature.
Overall, our results indicate that the transition is most likely second-order. Fitting the tetragonal distortion $(c-a)/a$
to the function $(T_\textup{c}-T)^\alpha$, as shown in  Figs.~\ref{fig:RPA_vs_SCAN}(c) and (f),
for MLFF-SCAN and MLFF-RPA$^\Delta$, respectively, yields
a transition temperature of 2585 K and 2546 K. Both are in very good agreement with the experimental value (2570 K~\cite{https://doi.org/10.1111/j.1151-2916.1985.tb15247.x}).
Again, MLFF-SCAN$^\Delta$ predicts a similar value of $T_c$ as MLFF-SCAN for the tetragonal to cubic phase transition [see Fig.~\ref{fig:RPA_vs_SCAN}(c)].

In summary, we have demonstrated the power of a combined approach of on-the-fly active learning and $\Delta$-ML.
Through rank compression of the local structures, we have reduced the number of high-level
quantum mechanical calculations to a very manageable level of just 168 medium-sized structures.
With less than 150 000 CPU hours, the final training at the RPA level was very affordable.
In fact, we could have halved the number of RPA calculations and obtained similar results.
Moreover, the present results again clearly demonstrate that, for solids, the RPA provides predictions
on par with the best experimental estimates for finite temperature properties, including structure predictions and phase transition temperatures.
The best available density functional SCAN is close, but compared to experiment the errors are somewhat larger.

The present work documents a major leap in the prediction of materials properties based on first principles.
Using machine-learned force fields, one can routinely predict finite temperature materials properties with DFT accuracy
at a fraction of the computational cost that would be required using standard DFT calculations.
Our present work shows that this leap also applies to high-accuracy many-body techniques.
Combining them with machine-learned force fields leads to unprecedented accuracy
and speed;  a new golden age for materials property predictions is dawning.

\begin{acknowledgments}
P. Liu thanks M. Kaltak for useful discussions.
This work was funded by the Advanced Materials Simulation Engineering Tool (AMSET) project,
sponsored by the US Naval Nuclear Laboratory (NNL) and directed by Materials Design, Inc.
The funding from the Austrian Science Fund (FWF) within the SFB TACO  (Grant No. F 81-N) is gratefully acknowledged.
\end{acknowledgments}

\bibliographystyle{apsrev4-1}
\bibliography{Reference} 

\begin{thebibliography}{38}%
\makeatletter
\providecommand \@ifxundefined [1]{%
 \@ifx{#1\undefined}
}%
\providecommand \@ifnum [1]{%
 \ifnum #1\expandafter \@firstoftwo
 \else \expandafter \@secondoftwo
 \fi
}%
\providecommand \@ifx [1]{%
 \ifx #1\expandafter \@firstoftwo
 \else \expandafter \@secondoftwo
 \fi
}%
\providecommand \natexlab [1]{#1}%
\providecommand \enquote  [1]{``#1''}%
\providecommand \bibnamefont  [1]{#1}%
\providecommand \bibfnamefont [1]{#1}%
\providecommand \citenamefont [1]{#1}%
\providecommand \href@noop [0]{\@secondoftwo}%
\providecommand \href [0]{\begingroup \@sanitize@url \@href}%
\providecommand \@href[1]{\@@startlink{#1}\@@href}%
\providecommand \@@href[1]{\endgroup#1\@@endlink}%
\providecommand \@sanitize@url [0]{\catcode `\\12\catcode `\$12\catcode
  `\&12\catcode `\#12\catcode `\^12\catcode `\_12\catcode `\%12\relax}%
\providecommand \@@startlink[1]{}%
\providecommand \@@endlink[0]{}%
\providecommand \url  [0]{\begingroup\@sanitize@url \@url }%
\providecommand \@url [1]{\endgroup\@href {#1}{\urlprefix }}%
\providecommand \urlprefix  [0]{URL }%
\providecommand \Eprint [0]{\href }%
\providecommand \doibase [0]{http://dx.doi.org/}%
\providecommand \selectlanguage [0]{\@gobble}%
\providecommand \bibinfo  [0]{\@secondoftwo}%
\providecommand \bibfield  [0]{\@secondoftwo}%
\providecommand \translation [1]{[#1]}%
\providecommand \BibitemOpen [0]{}%
\providecommand \bibitemStop [0]{}%
\providecommand \bibitemNoStop [0]{.\EOS\space}%
\providecommand \EOS [0]{\spacefactor3000\relax}%
\providecommand \BibitemShut  [1]{\csname bibitem#1\endcsname}%
\let\auto@bib@innerbib\@empty
\bibitem [{\citenamefont {Behler}\ and\ \citenamefont
  {Parrinello}(2007)}]{PhysRevLett.98.146401}%
  \BibitemOpen
  \bibfield  {author} {\bibinfo {author} {\bibfnamefont {J.}~\bibnamefont
  {Behler}}\ and\ \bibinfo {author} {\bibfnamefont {M.}~\bibnamefont
  {Parrinello}},\ }\href {\doibase 10.1103/PhysRevLett.98.146401} {\bibfield
  {journal} {\bibinfo  {journal} {Phys. Rev. Lett.}\ }\textbf {\bibinfo
  {volume} {98}},\ \bibinfo {pages} {146401} (\bibinfo {year}
  {2007})}\BibitemShut {NoStop}%
\bibitem [{\citenamefont {Bart\'ok}\ \emph {et~al.}(2010)\citenamefont
  {Bart\'ok}, \citenamefont {Payne}, \citenamefont {Kondor},\ and\
  \citenamefont {Cs\'anyi}}]{PhysRevLett.104.136403}%
  \BibitemOpen
  \bibfield  {author} {\bibinfo {author} {\bibfnamefont {A.~P.}\ \bibnamefont
  {Bart\'ok}}, \bibinfo {author} {\bibfnamefont {M.~C.}\ \bibnamefont {Payne}},
  \bibinfo {author} {\bibfnamefont {R.}~\bibnamefont {Kondor}}, \ and\ \bibinfo
  {author} {\bibfnamefont {G.}~\bibnamefont {Cs\'anyi}},\ }\href {\doibase
  10.1103/PhysRevLett.104.136403} {\bibfield  {journal} {\bibinfo  {journal}
  {Phys. Rev. Lett.}\ }\textbf {\bibinfo {volume} {104}},\ \bibinfo {pages}
  {136403} (\bibinfo {year} {2010})}\BibitemShut {NoStop}%
\bibitem [{\citenamefont {Botu}\ \emph {et~al.}(2017)\citenamefont {Botu},
  \citenamefont {Batra}, \citenamefont {Chapman},\ and\ \citenamefont
  {Ramprasad}}]{Botu_ML_review_JPCC2017}%
  \BibitemOpen
  \bibfield  {author} {\bibinfo {author} {\bibfnamefont {V.}~\bibnamefont
  {Botu}}, \bibinfo {author} {\bibfnamefont {R.}~\bibnamefont {Batra}},
  \bibinfo {author} {\bibfnamefont {J.}~\bibnamefont {Chapman}}, \ and\
  \bibinfo {author} {\bibfnamefont {R.}~\bibnamefont {Ramprasad}},\ }\href
  {\doibase 10.1021/acs.jpcc.6b10908} {\bibfield  {journal} {\bibinfo
  {journal} {J. Phys. Chem. C}\ }\textbf {\bibinfo {volume} {121}},\ \bibinfo
  {pages} {511} (\bibinfo {year} {2017})}\BibitemShut {NoStop}%
\bibitem [{\citenamefont {Behler}(2017)}]{Behler_NN_review2017}%
  \BibitemOpen
  \bibfield  {author} {\bibinfo {author} {\bibfnamefont {J.}~\bibnamefont
  {Behler}},\ }\href {\doibase 10.1002/anie.201703114} {\bibfield  {journal}
  {\bibinfo  {journal} {Angew. Chem. Int. Ed.}\ }\textbf {\bibinfo {volume}
  {56}},\ \bibinfo {pages} {12828} (\bibinfo {year} {2017})}\BibitemShut
  {NoStop}%
\bibitem [{\citenamefont {Bart\'ok}\ \emph {et~al.}(2018)\citenamefont
  {Bart\'ok}, \citenamefont {Kermode}, \citenamefont {Bernstein},\ and\
  \citenamefont {Cs\'anyi}}]{PhysRevX.8.041048}%
  \BibitemOpen
  \bibfield  {author} {\bibinfo {author} {\bibfnamefont {A.~P.}\ \bibnamefont
  {Bart\'ok}}, \bibinfo {author} {\bibfnamefont {J.}~\bibnamefont {Kermode}},
  \bibinfo {author} {\bibfnamefont {N.}~\bibnamefont {Bernstein}}, \ and\
  \bibinfo {author} {\bibfnamefont {G.}~\bibnamefont {Cs\'anyi}},\ }\href
  {\doibase 10.1103/PhysRevX.8.041048} {\bibfield  {journal} {\bibinfo
  {journal} {Phys. Rev. X}\ }\textbf {\bibinfo {volume} {8}},\ \bibinfo {pages}
  {041048} (\bibinfo {year} {2018})}\BibitemShut {NoStop}%
\bibitem [{\citenamefont {Seko}\ \emph {et~al.}(2014)\citenamefont {Seko},
  \citenamefont {Takahashi},\ and\ \citenamefont
  {Tanaka}}]{PhysRevB.90.024101}%
  \BibitemOpen
  \bibfield  {author} {\bibinfo {author} {\bibfnamefont {A.}~\bibnamefont
  {Seko}}, \bibinfo {author} {\bibfnamefont {A.}~\bibnamefont {Takahashi}}, \
  and\ \bibinfo {author} {\bibfnamefont {I.}~\bibnamefont {Tanaka}},\ }\href
  {\doibase 10.1103/PhysRevB.90.024101} {\bibfield  {journal} {\bibinfo
  {journal} {Phys. Rev. B}\ }\textbf {\bibinfo {volume} {90}},\ \bibinfo
  {pages} {024101} (\bibinfo {year} {2014})}\BibitemShut {NoStop}%
\bibitem [{\citenamefont {Shapeev}(2016)}]{Shapeev_MTP2016}%
  \BibitemOpen
  \bibfield  {author} {\bibinfo {author} {\bibfnamefont {A.~V.}\ \bibnamefont
  {Shapeev}},\ }\href {\doibase 10.1137/15M1054183} {\bibfield  {journal}
  {\bibinfo  {journal} {Multiscale Modeling \& Simulation}\ }\textbf {\bibinfo
  {volume} {14}},\ \bibinfo {pages} {1153} (\bibinfo {year}
  {2016})}\BibitemShut {NoStop}%
\bibitem [{\citenamefont {Glielmo}\ \emph {et~al.}(2018)\citenamefont
  {Glielmo}, \citenamefont {Zeni},\ and\ \citenamefont
  {De~Vita}}]{PhysRevB.97.184307}%
  \BibitemOpen
  \bibfield  {author} {\bibinfo {author} {\bibfnamefont {A.}~\bibnamefont
  {Glielmo}}, \bibinfo {author} {\bibfnamefont {C.}~\bibnamefont {Zeni}}, \
  and\ \bibinfo {author} {\bibfnamefont {A.}~\bibnamefont {De~Vita}},\ }\href
  {\doibase 10.1103/PhysRevB.97.184307} {\bibfield  {journal} {\bibinfo
  {journal} {Phys. Rev. B}\ }\textbf {\bibinfo {volume} {97}},\ \bibinfo
  {pages} {184307} (\bibinfo {year} {2018})}\BibitemShut {NoStop}%
\bibitem [{\citenamefont {Faber}\ \emph {et~al.}(2018)\citenamefont {Faber},
  \citenamefont {Christensen}, \citenamefont {Huang},\ and\ \citenamefont {von
  Lilienfeld}}]{doi:10.1063/1.5020710}%
  \BibitemOpen
  \bibfield  {author} {\bibinfo {author} {\bibfnamefont {F.~A.}\ \bibnamefont
  {Faber}}, \bibinfo {author} {\bibfnamefont {A.~S.}\ \bibnamefont
  {Christensen}}, \bibinfo {author} {\bibfnamefont {B.}~\bibnamefont {Huang}},
  \ and\ \bibinfo {author} {\bibfnamefont {O.~A.}\ \bibnamefont {von
  Lilienfeld}},\ }\href {\doibase 10.1063/1.5020710} {\bibfield  {journal}
  {\bibinfo  {journal} {J. Chem. Phys.}\ }\textbf {\bibinfo {volume} {148}},\
  \bibinfo {pages} {241717} (\bibinfo {year} {2018})}\BibitemShut {NoStop}%
\bibitem [{\citenamefont {De}\ \emph {et~al.}(2016)\citenamefont {De},
  \citenamefont {Bart\'{o}k}, \citenamefont {Cs\'{a}nyi},\ and\ \citenamefont
  {Ceriotti}}]{Michele_PCCP2016}%
  \BibitemOpen
  \bibfield  {author} {\bibinfo {author} {\bibfnamefont {S.}~\bibnamefont
  {De}}, \bibinfo {author} {\bibfnamefont {A.~P.}\ \bibnamefont {Bart\'{o}k}},
  \bibinfo {author} {\bibfnamefont {G.}~\bibnamefont {Cs\'{a}nyi}}, \ and\
  \bibinfo {author} {\bibfnamefont {M.}~\bibnamefont {Ceriotti}},\ }\href
  {\doibase 10.1039/C6CP00415F} {\bibfield  {journal} {\bibinfo  {journal}
  {Phys. Chem. Chem. Phys.}\ }\textbf {\bibinfo {volume} {18}},\ \bibinfo
  {pages} {13754} (\bibinfo {year} {2016})}\BibitemShut {NoStop}%
\bibitem [{\citenamefont {Schmidt}\ \emph {et~al.}(2019)\citenamefont
  {Schmidt}, \citenamefont {Marques}, \citenamefont {Botti},\ and\
  \citenamefont {Marques}}]{Marques_npj2019}%
  \BibitemOpen
  \bibfield  {author} {\bibinfo {author} {\bibfnamefont {J.}~\bibnamefont
  {Schmidt}}, \bibinfo {author} {\bibfnamefont {M.~R.~G.}\ \bibnamefont
  {Marques}}, \bibinfo {author} {\bibfnamefont {S.}~\bibnamefont {Botti}}, \
  and\ \bibinfo {author} {\bibfnamefont {M.~A.~L.}\ \bibnamefont {Marques}},\
  }\href {https://doi.org/10.1038/s41524-019-0221-0} {\bibfield  {journal}
  {\bibinfo  {journal} {npj Computational Materials}\ }\textbf {\bibinfo
  {volume} {5}},\ \bibinfo {pages} {83} (\bibinfo {year} {2019})}\BibitemShut
  {NoStop}%
\bibitem [{\citenamefont {Jinnouchi}\ and\ \citenamefont
  {Asahi}(2017)}]{doi:10.1021/acs.jpclett.7b02010}%
  \BibitemOpen
  \bibfield  {author} {\bibinfo {author} {\bibfnamefont {R.}~\bibnamefont
  {Jinnouchi}}\ and\ \bibinfo {author} {\bibfnamefont {R.}~\bibnamefont
  {Asahi}},\ }\href {\doibase 10.1021/acs.jpclett.7b02010} {\bibfield
  {journal} {\bibinfo  {journal} {J. Phys. Chem. Lett.}\ }\textbf {\bibinfo
  {volume} {8}},\ \bibinfo {pages} {4279} (\bibinfo {year} {2017})}\BibitemShut
  {NoStop}%
\bibitem [{\citenamefont {Li}\ \emph {et~al.}(2015)\citenamefont {Li},
  \citenamefont {Kermode},\ and\ \citenamefont
  {De~Vita}}]{PhysRevLett.114.096405}%
  \BibitemOpen
  \bibfield  {author} {\bibinfo {author} {\bibfnamefont {Z.}~\bibnamefont
  {Li}}, \bibinfo {author} {\bibfnamefont {J.~R.}\ \bibnamefont {Kermode}}, \
  and\ \bibinfo {author} {\bibfnamefont {A.}~\bibnamefont {De~Vita}},\ }\href
  {\doibase 10.1103/PhysRevLett.114.096405} {\bibfield  {journal} {\bibinfo
  {journal} {Phys. Rev. Lett.}\ }\textbf {\bibinfo {volume} {114}},\ \bibinfo
  {pages} {096405} (\bibinfo {year} {2015})}\BibitemShut {NoStop}%
\bibitem [{\citenamefont {Jacobsen}\ \emph {et~al.}(2018)\citenamefont
  {Jacobsen}, \citenamefont {J\o{}rgensen},\ and\ \citenamefont
  {Hammer}}]{PhysRevLett.120.026102}%
  \BibitemOpen
  \bibfield  {author} {\bibinfo {author} {\bibfnamefont {T.~L.}\ \bibnamefont
  {Jacobsen}}, \bibinfo {author} {\bibfnamefont {M.~S.}\ \bibnamefont
  {J\o{}rgensen}}, \ and\ \bibinfo {author} {\bibfnamefont {B.}~\bibnamefont
  {Hammer}},\ }\href {\doibase 10.1103/PhysRevLett.120.026102} {\bibfield
  {journal} {\bibinfo  {journal} {Phys. Rev. Lett.}\ }\textbf {\bibinfo
  {volume} {120}},\ \bibinfo {pages} {026102} (\bibinfo {year}
  {2018})}\BibitemShut {NoStop}%
\bibitem [{\citenamefont {Jinnouchi}\ \emph
  {et~al.}(2019{\natexlab{a}})\citenamefont {Jinnouchi}, \citenamefont
  {Lahnsteiner}, \citenamefont {Karsai}, \citenamefont {Kresse},\ and\
  \citenamefont {Bokdam}}]{PhysRevLett.122.225701}%
  \BibitemOpen
  \bibfield  {author} {\bibinfo {author} {\bibfnamefont {R.}~\bibnamefont
  {Jinnouchi}}, \bibinfo {author} {\bibfnamefont {J.}~\bibnamefont
  {Lahnsteiner}}, \bibinfo {author} {\bibfnamefont {F.}~\bibnamefont {Karsai}},
  \bibinfo {author} {\bibfnamefont {G.}~\bibnamefont {Kresse}}, \ and\ \bibinfo
  {author} {\bibfnamefont {M.}~\bibnamefont {Bokdam}},\ }\href {\doibase
  10.1103/PhysRevLett.122.225701} {\bibfield  {journal} {\bibinfo  {journal}
  {Phys. Rev. Lett.}\ }\textbf {\bibinfo {volume} {122}},\ \bibinfo {pages}
  {225701} (\bibinfo {year} {2019}{\natexlab{a}})}\BibitemShut {NoStop}%
\bibitem [{\citenamefont {Ramakrishnan}\ \emph {et~al.}(2015)\citenamefont
  {Ramakrishnan}, \citenamefont {Dral}, \citenamefont {Rupp},\ and\
  \citenamefont {von Lilienfeld}}]{Anatole_2015JCTC}%
  \BibitemOpen
  \bibfield  {author} {\bibinfo {author} {\bibfnamefont {R.}~\bibnamefont
  {Ramakrishnan}}, \bibinfo {author} {\bibfnamefont {P.~O.}\ \bibnamefont
  {Dral}}, \bibinfo {author} {\bibfnamefont {M.}~\bibnamefont {Rupp}}, \ and\
  \bibinfo {author} {\bibfnamefont {O.~A.}\ \bibnamefont {von Lilienfeld}},\
  }\href {\doibase 10.1021/acs.jctc.5b00099} {\bibfield  {journal} {\bibinfo
  {journal} {J. Chem. Theory Comput.}\ }\textbf {\bibinfo {volume} {11}},\
  \bibinfo {pages} {2087} (\bibinfo {year} {2015})}\BibitemShut {NoStop}%
\bibitem [{\citenamefont {Bart{\'o}k}\ \emph {et~al.}(2017)\citenamefont
  {Bart{\'o}k}, \citenamefont {De}, \citenamefont {Poelking}, \citenamefont
  {Bernstein}, \citenamefont {Kermode}, \citenamefont {Cs{\'a}nyi},\ and\
  \citenamefont {Ceriotti}}]{Bartokek_SA2017}%
  \BibitemOpen
  \bibfield  {author} {\bibinfo {author} {\bibfnamefont {A.~P.}\ \bibnamefont
  {Bart{\'o}k}}, \bibinfo {author} {\bibfnamefont {S.}~\bibnamefont {De}},
  \bibinfo {author} {\bibfnamefont {C.}~\bibnamefont {Poelking}}, \bibinfo
  {author} {\bibfnamefont {N.}~\bibnamefont {Bernstein}}, \bibinfo {author}
  {\bibfnamefont {J.~R.}\ \bibnamefont {Kermode}}, \bibinfo {author}
  {\bibfnamefont {G.}~\bibnamefont {Cs{\'a}nyi}}, \ and\ \bibinfo {author}
  {\bibfnamefont {M.}~\bibnamefont {Ceriotti}},\ }\href
  {https://advances.sciencemag.org/content/3/12/e1701816} {\bibfield  {journal}
  {\bibinfo  {journal} {Science Advances}\ }\textbf {\bibinfo {volume} {3}},\
  \bibinfo {pages} {e1701816} (\bibinfo {year} {2017})}\BibitemShut {NoStop}%
\bibitem [{\citenamefont {Chmiela}\ \emph {et~al.}(2018)\citenamefont
  {Chmiela}, \citenamefont {Sauceda}, \citenamefont {M\"{u}ller},\ and\
  \citenamefont {Tkatchenko}}]{Chmiela2018}%
  \BibitemOpen
  \bibfield  {author} {\bibinfo {author} {\bibfnamefont {S.}~\bibnamefont
  {Chmiela}}, \bibinfo {author} {\bibfnamefont {H.~E.}\ \bibnamefont
  {Sauceda}}, \bibinfo {author} {\bibfnamefont {K.-R.}\ \bibnamefont
  {M\"{u}ller}}, \ and\ \bibinfo {author} {\bibfnamefont {A.}~\bibnamefont
  {Tkatchenko}},\ }\href {\doibase 10.1038/s41467-018-06169-2} {\bibfield
  {journal} {\bibinfo  {journal} {Nat. Commun.}\ }\textbf {\bibinfo {volume}
  {9}},\ \bibinfo {pages} {3887} (\bibinfo {year} {2018})}\BibitemShut
  {NoStop}%
\bibitem [{\citenamefont {Sauceda}\ \emph {et~al.}(2019)\citenamefont
  {Sauceda}, \citenamefont {Chmiela}, \citenamefont {Poltavsky}, \citenamefont
  {M\"{u}ller},\ and\ \citenamefont {Tkatchenko}}]{doi:10.1063/1.5078687}%
  \BibitemOpen
  \bibfield  {author} {\bibinfo {author} {\bibfnamefont {H.~E.}\ \bibnamefont
  {Sauceda}}, \bibinfo {author} {\bibfnamefont {S.}~\bibnamefont {Chmiela}},
  \bibinfo {author} {\bibfnamefont {I.}~\bibnamefont {Poltavsky}}, \bibinfo
  {author} {\bibfnamefont {K.-R.}\ \bibnamefont {M\"{u}ller}}, \ and\ \bibinfo
  {author} {\bibfnamefont {A.}~\bibnamefont {Tkatchenko}},\ }\href {\doibase
  10.1063/1.5078687} {\bibfield  {journal} {\bibinfo  {journal} {J. Chem.
  Phys.}\ }\textbf {\bibinfo {volume} {150}},\ \bibinfo {pages} {114102}
  (\bibinfo {year} {2019})}\BibitemShut {NoStop}%
\bibitem [{\citenamefont {Smith}\ \emph {et~al.}(2019)\citenamefont {Smith},
  \citenamefont {Nebgen}, \citenamefont {Zubatyuk}, \citenamefont {Lubbers},
  \citenamefont {Devereux}, \citenamefont {Barros}, \citenamefont {Tretiak},
  \citenamefont {Isayev},\ and\ \citenamefont {Roitberg}}]{Smith2019}%
  \BibitemOpen
  \bibfield  {author} {\bibinfo {author} {\bibfnamefont {J.~S.}\ \bibnamefont
  {Smith}}, \bibinfo {author} {\bibfnamefont {B.~T.}\ \bibnamefont {Nebgen}},
  \bibinfo {author} {\bibfnamefont {R.}~\bibnamefont {Zubatyuk}}, \bibinfo
  {author} {\bibfnamefont {N.}~\bibnamefont {Lubbers}}, \bibinfo {author}
  {\bibfnamefont {C.}~\bibnamefont {Devereux}}, \bibinfo {author}
  {\bibfnamefont {K.}~\bibnamefont {Barros}}, \bibinfo {author} {\bibfnamefont
  {S.}~\bibnamefont {Tretiak}}, \bibinfo {author} {\bibfnamefont
  {O.}~\bibnamefont {Isayev}}, \ and\ \bibinfo {author} {\bibfnamefont {A.~E.}\
  \bibnamefont {Roitberg}},\ }\href {\doibase 10.1038/s41467-019-10827-4}
  {\bibfield  {journal} {\bibinfo  {journal} {Nat. Commun.}\ }\textbf {\bibinfo
  {volume} {10}},\ \bibinfo {pages} {2903} (\bibinfo {year}
  {2019})}\BibitemShut {NoStop}%
\bibitem [{\citenamefont {Bogojeski}\ \emph {et~al.}(2020)\citenamefont
  {Bogojeski}, \citenamefont {Vogt-Maranto}, \citenamefont {Tuckerman},
  \citenamefont {M\"{u}ller},\ and\ \citenamefont {Burke}}]{Bogojeski2020}%
  \BibitemOpen
  \bibfield  {author} {\bibinfo {author} {\bibfnamefont {M.}~\bibnamefont
  {Bogojeski}}, \bibinfo {author} {\bibfnamefont {L.}~\bibnamefont
  {Vogt-Maranto}}, \bibinfo {author} {\bibfnamefont {M.~E.}\ \bibnamefont
  {Tuckerman}}, \bibinfo {author} {\bibfnamefont {K.-R.}\ \bibnamefont
  {M\"{u}ller}}, \ and\ \bibinfo {author} {\bibfnamefont {K.}~\bibnamefont
  {Burke}},\ }\href {\doibase 10.1038/s41467-020-19093-1} {\bibfield  {journal}
  {\bibinfo  {journal} {Nat. Commun.}\ }\textbf {\bibinfo {volume} {11}},\
  \bibinfo {pages} {5223} (\bibinfo {year} {2020})}\BibitemShut {NoStop}%
\bibitem [{\citenamefont {Chehaibou}\ \emph {et~al.}(2019)\citenamefont
  {Chehaibou}, \citenamefont {Badawi}, \citenamefont {Bu\v{c}ko},\ and\
  \citenamefont {Rocca}}]{MLTP_2019JCTC}%
  \BibitemOpen
  \bibfield  {author} {\bibinfo {author} {\bibfnamefont {B.}~\bibnamefont
  {Chehaibou}}, \bibinfo {author} {\bibfnamefont {M.}~\bibnamefont {Badawi}},
  \bibinfo {author} {\bibfnamefont {T.}~\bibnamefont {Bu\v{c}ko}, \bibfnamefont
  {Tom\'{a}\v{s}and~Bazhirov}}, \ and\ \bibinfo {author} {\bibfnamefont
  {D.}~\bibnamefont {Rocca}},\ }\href {\doibase 10.1021/acs.jctc.9b00782}
  {\bibfield  {journal} {\bibinfo  {journal} {J. Chem. Theory Comput.}\
  }\textbf {\bibinfo {volume} {15}},\ \bibinfo {pages} {6333} (\bibinfo {year}
  {2019})}\BibitemShut {NoStop}%
\bibitem [{\citenamefont {Deringer}\ \emph {et~al.}(2021)\citenamefont
  {Deringer}, \citenamefont {Bernstein}, \citenamefont {Cs\'anyi},
  \citenamefont {Ben~Mahmoud}, \citenamefont {Ceriotti}, \citenamefont
  {Wilson}, \citenamefont {Drabold},\ and\ \citenamefont
  {Elliott}}]{Deringer2021}%
  \BibitemOpen
  \bibfield  {author} {\bibinfo {author} {\bibfnamefont {V.~L.}\ \bibnamefont
  {Deringer}}, \bibinfo {author} {\bibfnamefont {N.}~\bibnamefont {Bernstein}},
  \bibinfo {author} {\bibfnamefont {G.}~\bibnamefont {Cs\'anyi}}, \bibinfo
  {author} {\bibfnamefont {C.}~\bibnamefont {Ben~Mahmoud}}, \bibinfo {author}
  {\bibfnamefont {M.}~\bibnamefont {Ceriotti}}, \bibinfo {author}
  {\bibfnamefont {M.}~\bibnamefont {Wilson}}, \bibinfo {author} {\bibfnamefont
  {D.~A.}\ \bibnamefont {Drabold}}, \ and\ \bibinfo {author} {\bibfnamefont
  {S.~R.}\ \bibnamefont {Elliott}},\ }\href {\doibase
  10.1038/s41586-020-03072-z} {\bibfield  {journal} {\bibinfo  {journal}
  {Nature}\ }\textbf {\bibinfo {volume} {589}},\ \bibinfo {pages} {59}
  (\bibinfo {year} {2021})}\BibitemShut {NoStop}%
\bibitem [{\citenamefont {Jinnouchi}\ \emph
  {et~al.}(2019{\natexlab{b}})\citenamefont {Jinnouchi}, \citenamefont
  {Karsai},\ and\ \citenamefont {Kresse}}]{PhysRevB.100.014105}%
  \BibitemOpen
  \bibfield  {author} {\bibinfo {author} {\bibfnamefont {R.}~\bibnamefont
  {Jinnouchi}}, \bibinfo {author} {\bibfnamefont {F.}~\bibnamefont {Karsai}}, \
  and\ \bibinfo {author} {\bibfnamefont {G.}~\bibnamefont {Kresse}},\ }\href
  {\doibase 10.1103/PhysRevB.100.014105} {\bibfield  {journal} {\bibinfo
  {journal} {Phys. Rev. B}\ }\textbf {\bibinfo {volume} {100}},\ \bibinfo
  {pages} {014105} (\bibinfo {year} {2019}{\natexlab{b}})}\BibitemShut
  {NoStop}%
\bibitem [{\citenamefont {Verdi}\ \emph {et~al.}(2021)\citenamefont {Verdi},
  \citenamefont {Karsai}, \citenamefont {Liu}, \citenamefont {Jinnouchi},\ and\
  \citenamefont {Kresse}}]{Carla_2021}%
  \BibitemOpen
  \bibfield  {author} {\bibinfo {author} {\bibfnamefont {C.}~\bibnamefont
  {Verdi}}, \bibinfo {author} {\bibfnamefont {F.}~\bibnamefont {Karsai}},
  \bibinfo {author} {\bibfnamefont {P.}~\bibnamefont {Liu}}, \bibinfo {author}
  {\bibfnamefont {R.}~\bibnamefont {Jinnouchi}}, \ and\ \bibinfo {author}
  {\bibfnamefont {G.}~\bibnamefont {Kresse}},\ }\href@noop {} {\  (\bibinfo
  {year} {npj Comput. Mater., accepted, 2021})}\BibitemShut {NoStop}%
\bibitem [{\citenamefont {Perdew}\ \emph {et~al.}(1996)\citenamefont {Perdew},
  \citenamefont {Burke},\ and\ \citenamefont
  {Ernzerhof}}]{PhysRevLett.77.3865}%
  \BibitemOpen
  \bibfield  {author} {\bibinfo {author} {\bibfnamefont {J.~P.}\ \bibnamefont
  {Perdew}}, \bibinfo {author} {\bibfnamefont {K.}~\bibnamefont {Burke}}, \
  and\ \bibinfo {author} {\bibfnamefont {M.}~\bibnamefont {Ernzerhof}},\ }\href
  {\doibase 10.1103/PhysRevLett.77.3865} {\bibfield  {journal} {\bibinfo
  {journal} {Phys. Rev. Lett.}\ }\textbf {\bibinfo {volume} {77}},\ \bibinfo
  {pages} {3865} (\bibinfo {year} {1996})}\BibitemShut {NoStop}%
\bibitem [{SM()}]{SM}%
  \BibitemOpen
  \href@noop {} {\emph {\bibinfo {title} {\rm See Supplemental Material for the
  details of first-principles calculations, MLFF training and validation, phase
  transitions of zirconia within the QHA as well as singular value
  decomposition rank compression.}}}\BibitemShut {Stop}%
\bibitem [{\citenamefont {Mahoney}\ and\ \citenamefont
  {Drineas}(2009)}]{Mahoney697}%
  \BibitemOpen
  \bibfield  {author} {\bibinfo {author} {\bibfnamefont {M.~W.}\ \bibnamefont
  {Mahoney}}\ and\ \bibinfo {author} {\bibfnamefont {P.}~\bibnamefont
  {Drineas}},\ }\href {\doibase 10.1073/pnas.0803205106} {\bibfield  {journal}
  {\bibinfo  {journal} {Proc. Natl. Acad. Sci. U.S.A.}\ }\textbf {\bibinfo
  {volume} {106}},\ \bibinfo {pages} {697} (\bibinfo {year}
  {2009})}\BibitemShut {NoStop}%
\bibitem [{\citenamefont {Sun}\ \emph {et~al.}(2015)\citenamefont {Sun},
  \citenamefont {Ruzsinszky},\ and\ \citenamefont
  {Perdew}}]{PhysRevLett.115.036402}%
  \BibitemOpen
  \bibfield  {author} {\bibinfo {author} {\bibfnamefont {J.}~\bibnamefont
  {Sun}}, \bibinfo {author} {\bibfnamefont {A.}~\bibnamefont {Ruzsinszky}}, \
  and\ \bibinfo {author} {\bibfnamefont {J.~P.}\ \bibnamefont {Perdew}},\
  }\href {\doibase 10.1103/PhysRevLett.115.036402} {\bibfield  {journal}
  {\bibinfo  {journal} {Phys. Rev. Lett.}\ }\textbf {\bibinfo {volume} {115}},\
  \bibinfo {pages} {036402} (\bibinfo {year} {2015})}\BibitemShut {NoStop}%
\bibitem [{\citenamefont {Kaltak}\ \emph {et~al.}(2014)\citenamefont {Kaltak},
  \citenamefont {Klime\v{s}},\ and\ \citenamefont
  {Kresse}}]{PhysRevB.90.054115}%
  \BibitemOpen
  \bibfield  {author} {\bibinfo {author} {\bibfnamefont {M.}~\bibnamefont
  {Kaltak}}, \bibinfo {author} {\bibfnamefont {J.}~\bibnamefont {Klime\v{s}}},
  \ and\ \bibinfo {author} {\bibfnamefont {G.}~\bibnamefont {Kresse}},\ }\href
  {\doibase 10.1103/PhysRevB.90.054115} {\bibfield  {journal} {\bibinfo
  {journal} {Phys. Rev. B}\ }\textbf {\bibinfo {volume} {90}},\ \bibinfo
  {pages} {054115} (\bibinfo {year} {2014})}\BibitemShut {NoStop}%
\bibitem [{\citenamefont {Ramberger}\ \emph {et~al.}(2017)\citenamefont
  {Ramberger}, \citenamefont {Sch\"afer},\ and\ \citenamefont
  {Kresse}}]{PhysRevLett.118.106403}%
  \BibitemOpen
  \bibfield  {author} {\bibinfo {author} {\bibfnamefont {B.}~\bibnamefont
  {Ramberger}}, \bibinfo {author} {\bibfnamefont {T.}~\bibnamefont
  {Sch\"afer}}, \ and\ \bibinfo {author} {\bibfnamefont {G.}~\bibnamefont
  {Kresse}},\ }\href {\doibase 10.1103/PhysRevLett.118.106403} {\bibfield
  {journal} {\bibinfo  {journal} {Phys. Rev. Lett.}\ }\textbf {\bibinfo
  {volume} {118}},\ \bibinfo {pages} {106403} (\bibinfo {year}
  {2017})}\BibitemShut {NoStop}%
\bibitem [{\citenamefont {Jia}\ \emph {et~al.}(2019)\citenamefont {Jia},
  \citenamefont {Kresse}, \citenamefont {Franchini}, \citenamefont {Liu},
  \citenamefont {Wang}, \citenamefont {Stroppa},\ and\ \citenamefont
  {Ren}}]{PhysRevMaterials.3.103801}%
  \BibitemOpen
  \bibfield  {author} {\bibinfo {author} {\bibfnamefont {F.}~\bibnamefont
  {Jia}}, \bibinfo {author} {\bibfnamefont {G.}~\bibnamefont {Kresse}},
  \bibinfo {author} {\bibfnamefont {C.}~\bibnamefont {Franchini}}, \bibinfo
  {author} {\bibfnamefont {P.}~\bibnamefont {Liu}}, \bibinfo {author}
  {\bibfnamefont {J.}~\bibnamefont {Wang}}, \bibinfo {author} {\bibfnamefont
  {A.}~\bibnamefont {Stroppa}}, \ and\ \bibinfo {author} {\bibfnamefont
  {W.}~\bibnamefont {Ren}},\ }\href {\doibase
  10.1103/PhysRevMaterials.3.103801} {\bibfield  {journal} {\bibinfo  {journal}
  {Phys. Rev. Materials}\ }\textbf {\bibinfo {volume} {3}},\ \bibinfo {pages}
  {103801} (\bibinfo {year} {2019})}\BibitemShut {NoStop}%
\bibitem [{\citenamefont {Mayr-Schm\"olzer}\ \emph {et~al.}(2020)\citenamefont
  {Mayr-Schm\"olzer}, \citenamefont {Planer}, \citenamefont {Redinger},
  \citenamefont {Gr\"uneis},\ and\ \citenamefont
  {Mittendorfer}}]{PhysRevResearch.2.043361}%
  \BibitemOpen
  \bibfield  {author} {\bibinfo {author} {\bibfnamefont {W.}~\bibnamefont
  {Mayr-Schm\"olzer}}, \bibinfo {author} {\bibfnamefont {J.}~\bibnamefont
  {Planer}}, \bibinfo {author} {\bibfnamefont {J.}~\bibnamefont {Redinger}},
  \bibinfo {author} {\bibfnamefont {A.}~\bibnamefont {Gr\"uneis}}, \ and\
  \bibinfo {author} {\bibfnamefont {F.}~\bibnamefont {Mittendorfer}},\ }\href
  {\doibase 10.1103/PhysRevResearch.2.043361} {\bibfield  {journal} {\bibinfo
  {journal} {Phys. Rev. Research}\ }\textbf {\bibinfo {volume} {2}},\ \bibinfo
  {pages} {043361} (\bibinfo {year} {2020})}\BibitemShut {NoStop}%
\bibitem [{\citenamefont {Kisi}\ and\ \citenamefont {Howard}(1998)}]{kisi1998}%
  \BibitemOpen
  \bibfield  {author} {\bibinfo {author} {\bibfnamefont {E.~H.}\ \bibnamefont
  {Kisi}}\ and\ \bibinfo {author} {\bibfnamefont {C.}~\bibnamefont {Howard}},\
  }in\ \href {\doibase 10.4028/www.scientific.net/KEM.153-154.1} {\emph
  {\bibinfo {booktitle} {Zirconia Engineering Ceramics}}},\ \bibinfo {series}
  {Key Engineering Materials}, Vol.\ \bibinfo {volume} {153}\ (\bibinfo
  {publisher} {Trans Tech Publications Ltd},\ \bibinfo {year} {1998})\ pp.\
  \bibinfo {pages} {1--36}\BibitemShut {NoStop}%
\bibitem [{\citenamefont {Aldebert}\ and\ \citenamefont
  {Traverse}(1985)}]{https://doi.org/10.1111/j.1151-2916.1985.tb15247.x}%
  \BibitemOpen
  \bibfield  {author} {\bibinfo {author} {\bibfnamefont {P.}~\bibnamefont
  {Aldebert}}\ and\ \bibinfo {author} {\bibfnamefont {J.-P.}\ \bibnamefont
  {Traverse}},\ }\href {\doibase
  https://doi.org/10.1111/j.1151-2916.1985.tb15247.x} {\bibfield  {journal}
  {\bibinfo  {journal} {J. Am. Ceram. Soc.}\ }\textbf {\bibinfo {volume}
  {68}},\ \bibinfo {pages} {34} (\bibinfo {year} {1985})}\BibitemShut {NoStop}%
\bibitem [{\citenamefont {Moriya}\ and\ \citenamefont
  {Navrotsky}(2006)}]{MORIYA2006211}%
  \BibitemOpen
  \bibfield  {author} {\bibinfo {author} {\bibfnamefont {Y.}~\bibnamefont
  {Moriya}}\ and\ \bibinfo {author} {\bibfnamefont {A.}~\bibnamefont
  {Navrotsky}},\ }\href {\doibase https://doi.org/10.1016/j.jct.2005.05.002}
  {\bibfield  {journal} {\bibinfo  {journal} {The Journal of Chemical
  Thermodynamics}\ }\textbf {\bibinfo {volume} {38}},\ \bibinfo {pages} {211}
  (\bibinfo {year} {2006})}\BibitemShut {NoStop}%
\bibitem [{\citenamefont {Stefanovich}\ \emph {et~al.}(1994)\citenamefont
  {Stefanovich}, \citenamefont {Shluger},\ and\ \citenamefont
  {Catlow}}]{PhysRevB.49.11560}%
  \BibitemOpen
  \bibfield  {author} {\bibinfo {author} {\bibfnamefont {E.~V.}\ \bibnamefont
  {Stefanovich}}, \bibinfo {author} {\bibfnamefont {A.~L.}\ \bibnamefont
  {Shluger}}, \ and\ \bibinfo {author} {\bibfnamefont {C.~R.~A.}\ \bibnamefont
  {Catlow}},\ }\href {\doibase 10.1103/PhysRevB.49.11560} {\bibfield  {journal}
  {\bibinfo  {journal} {Phys. Rev. B}\ }\textbf {\bibinfo {volume} {49}},\
  \bibinfo {pages} {11560} (\bibinfo {year} {1994})}\BibitemShut {NoStop}%
\bibitem [{\citenamefont {Liu}\ \emph {et~al.}(2021)\citenamefont {Liu},
  \citenamefont {Verdi}, \citenamefont {Karsai},\ and\ \citenamefont
  {Kresse}}]{Peitao_2020}%
  \BibitemOpen
  \bibfield  {author} {\bibinfo {author} {\bibfnamefont {P.}~\bibnamefont
  {Liu}}, \bibinfo {author} {\bibfnamefont {C.}~\bibnamefont {Verdi}}, \bibinfo
  {author} {\bibfnamefont {F.}~\bibnamefont {Karsai}}, \ and\ \bibinfo {author}
  {\bibfnamefont {G.}~\bibnamefont {Kresse}},\ }\href {\doibase
  10.1103/PhysRevMaterials.5.053804} {\bibfield  {journal} {\bibinfo  {journal}
  {Phys. Rev. Materials}\ }\textbf {\bibinfo {volume} {5}},\ \bibinfo {pages}
  {053804} (\bibinfo {year} {2021})}\BibitemShut {NoStop}%
\end{thebibliography}%


\begin{thebibliography}{27}%
\makeatletter
\providecommand \@ifxundefined [1]{%
 \@ifx{#1\undefined}
}%
\providecommand \@ifnum [1]{%
 \ifnum #1\expandafter \@firstoftwo
 \else \expandafter \@secondoftwo
 \fi
}%
\providecommand \@ifx [1]{%
 \ifx #1\expandafter \@firstoftwo
 \else \expandafter \@secondoftwo
 \fi
}%
\providecommand \natexlab [1]{#1}%
\providecommand \enquote  [1]{``#1''}%
\providecommand \bibnamefont  [1]{#1}%
\providecommand \bibfnamefont [1]{#1}%
\providecommand \citenamefont [1]{#1}%
\providecommand \href@noop [0]{\@secondoftwo}%
\providecommand \href [0]{\begingroup \@sanitize@url \@href}%
\providecommand \@href[1]{\@@startlink{#1}\@@href}%
\providecommand \@@href[1]{\endgroup#1\@@endlink}%
\providecommand \@sanitize@url [0]{\catcode `\\12\catcode `\$12\catcode
  `\&12\catcode `\#12\catcode `\^12\catcode `\_12\catcode `\%12\relax}%
\providecommand \@@startlink[1]{}%
\providecommand \@@endlink[0]{}%
\providecommand \url  [0]{\begingroup\@sanitize@url \@url }%
\providecommand \@url [1]{\endgroup\@href {#1}{\urlprefix }}%
\providecommand \urlprefix  [0]{URL }%
\providecommand \Eprint [0]{\href }%
\providecommand \doibase [0]{http://dx.doi.org/}%
\providecommand \selectlanguage [0]{\@gobble}%
\providecommand \bibinfo  [0]{\@secondoftwo}%
\providecommand \bibfield  [0]{\@secondoftwo}%
\providecommand \translation [1]{[#1]}%
\providecommand \BibitemOpen [0]{}%
\providecommand \bibitemStop [0]{}%
\providecommand \bibitemNoStop [0]{.\EOS\space}%
\providecommand \EOS [0]{\spacefactor3000\relax}%
\providecommand \BibitemShut  [1]{\csname bibitem#1\endcsname}%
\let\auto@bib@innerbib\@empty
\bibitem [{\citenamefont {Bl\"ochl}(1994)}]{PhysRevB.50.17953}%
  \BibitemOpen
  \bibfield  {author} {\bibinfo {author} {\bibfnamefont {P.~E.}\ \bibnamefont
  {Bl\"ochl}},\ }\href {\doibase 10.1103/PhysRevB.50.17953} {\bibfield
  {journal} {\bibinfo  {journal} {Phys. Rev. B}\ }\textbf {\bibinfo {volume}
  {50}},\ \bibinfo {pages} {17953} (\bibinfo {year} {1994})}\BibitemShut
  {NoStop}%
\bibitem [{\citenamefont {Kresse}\ and\ \citenamefont
  {Hafner}(1993)}]{PhysRevB.47.558}%
  \BibitemOpen
  \bibfield  {author} {\bibinfo {author} {\bibfnamefont {G.}~\bibnamefont
  {Kresse}}\ and\ \bibinfo {author} {\bibfnamefont {J.}~\bibnamefont
  {Hafner}},\ }\href {\doibase 10.1103/PhysRevB.47.558} {\bibfield  {journal}
  {\bibinfo  {journal} {Phys. Rev. B}\ }\textbf {\bibinfo {volume} {47}},\
  \bibinfo {pages} {558} (\bibinfo {year} {1993})}\BibitemShut {NoStop}%
\bibitem [{\citenamefont {Kresse}\ and\ \citenamefont
  {Furthm\"uller}(1996)}]{PhysRevB.54.11169}%
  \BibitemOpen
  \bibfield  {author} {\bibinfo {author} {\bibfnamefont {G.}~\bibnamefont
  {Kresse}}\ and\ \bibinfo {author} {\bibfnamefont {J.}~\bibnamefont
  {Furthm\"uller}},\ }\href {\doibase 10.1103/PhysRevB.54.11169} {\bibfield
  {journal} {\bibinfo  {journal} {Phys. Rev. B}\ }\textbf {\bibinfo {volume}
  {54}},\ \bibinfo {pages} {11169} (\bibinfo {year} {1996})}\BibitemShut
  {NoStop}%
\bibitem [{\citenamefont {Perdew}\ \emph {et~al.}(1996)\citenamefont {Perdew},
  \citenamefont {Burke},\ and\ \citenamefont
  {Ernzerhof}}]{PhysRevLett.77.3865}%
  \BibitemOpen
  \bibfield  {author} {\bibinfo {author} {\bibfnamefont {J.~P.}\ \bibnamefont
  {Perdew}}, \bibinfo {author} {\bibfnamefont {K.}~\bibnamefont {Burke}}, \
  and\ \bibinfo {author} {\bibfnamefont {M.}~\bibnamefont {Ernzerhof}},\ }\href
  {\doibase 10.1103/PhysRevLett.77.3865} {\bibfield  {journal} {\bibinfo
  {journal} {Phys. Rev. Lett.}\ }\textbf {\bibinfo {volume} {77}},\ \bibinfo
  {pages} {3865} (\bibinfo {year} {1996})}\BibitemShut {NoStop}%
\bibitem [{\citenamefont {Sun}\ \emph {et~al.}(2015)\citenamefont {Sun},
  \citenamefont {Ruzsinszky},\ and\ \citenamefont
  {Perdew}}]{PhysRevLett.115.036402}%
  \BibitemOpen
  \bibfield  {author} {\bibinfo {author} {\bibfnamefont {J.}~\bibnamefont
  {Sun}}, \bibinfo {author} {\bibfnamefont {A.}~\bibnamefont {Ruzsinszky}}, \
  and\ \bibinfo {author} {\bibfnamefont {J.~P.}\ \bibnamefont {Perdew}},\
  }\href {\doibase 10.1103/PhysRevLett.115.036402} {\bibfield  {journal}
  {\bibinfo  {journal} {Phys. Rev. Lett.}\ }\textbf {\bibinfo {volume} {115}},\
  \bibinfo {pages} {036402} (\bibinfo {year} {2015})}\BibitemShut {NoStop}%
\bibitem [{\citenamefont {Togo}\ \emph {et~al.}(2008)\citenamefont {Togo},
  \citenamefont {Oba},\ and\ \citenamefont {Tanaka}}]{PhysRevB.78.134106}%
  \BibitemOpen
  \bibfield  {author} {\bibinfo {author} {\bibfnamefont {A.}~\bibnamefont
  {Togo}}, \bibinfo {author} {\bibfnamefont {F.}~\bibnamefont {Oba}}, \ and\
  \bibinfo {author} {\bibfnamefont {I.}~\bibnamefont {Tanaka}},\ }\href
  {\doibase 10.1103/PhysRevB.78.134106} {\bibfield  {journal} {\bibinfo
  {journal} {Phys. Rev. B}\ }\textbf {\bibinfo {volume} {78}},\ \bibinfo
  {pages} {134106} (\bibinfo {year} {2008})}\BibitemShut {NoStop}%
\bibitem [{\citenamefont {Gonze}\ and\ \citenamefont
  {Lee}(1997)}]{PhysRevB.55.10355}%
  \BibitemOpen
  \bibfield  {author} {\bibinfo {author} {\bibfnamefont {X.}~\bibnamefont
  {Gonze}}\ and\ \bibinfo {author} {\bibfnamefont {C.}~\bibnamefont {Lee}},\
  }\href {\doibase 10.1103/PhysRevB.55.10355} {\bibfield  {journal} {\bibinfo
  {journal} {Phys. Rev. B}\ }\textbf {\bibinfo {volume} {55}},\ \bibinfo
  {pages} {10355} (\bibinfo {year} {1997})}\BibitemShut {NoStop}%
\bibitem [{\citenamefont {Perdew}\ \emph {et~al.}(2008)\citenamefont {Perdew},
  \citenamefont {Ruzsinszky}, \citenamefont {Csonka}, \citenamefont {Vydrov},
  \citenamefont {Scuseria}, \citenamefont {Constantin}, \citenamefont {Zhou},\
  and\ \citenamefont {Burke}}]{PhysRevLett.100.136406}%
  \BibitemOpen
  \bibfield  {author} {\bibinfo {author} {\bibfnamefont {J.~P.}\ \bibnamefont
  {Perdew}}, \bibinfo {author} {\bibfnamefont {A.}~\bibnamefont {Ruzsinszky}},
  \bibinfo {author} {\bibfnamefont {G.~I.}\ \bibnamefont {Csonka}}, \bibinfo
  {author} {\bibfnamefont {O.~A.}\ \bibnamefont {Vydrov}}, \bibinfo {author}
  {\bibfnamefont {G.~E.}\ \bibnamefont {Scuseria}}, \bibinfo {author}
  {\bibfnamefont {L.~A.}\ \bibnamefont {Constantin}}, \bibinfo {author}
  {\bibfnamefont {X.}~\bibnamefont {Zhou}}, \ and\ \bibinfo {author}
  {\bibfnamefont {K.}~\bibnamefont {Burke}},\ }\href {\doibase
  10.1103/PhysRevLett.100.136406} {\bibfield  {journal} {\bibinfo  {journal}
  {Phys. Rev. Lett.}\ }\textbf {\bibinfo {volume} {100}},\ \bibinfo {pages}
  {136406} (\bibinfo {year} {2008})}\BibitemShut {NoStop}%
\bibitem [{\citenamefont {Mayr-Schm\"olzer}\ \emph {et~al.}(2020)\citenamefont
  {Mayr-Schm\"olzer}, \citenamefont {Planer}, \citenamefont {Redinger},
  \citenamefont {Gr\"uneis},\ and\ \citenamefont
  {Mittendorfer}}]{PhysRevResearch.2.043361}%
  \BibitemOpen
  \bibfield  {author} {\bibinfo {author} {\bibfnamefont {W.}~\bibnamefont
  {Mayr-Schm\"olzer}}, \bibinfo {author} {\bibfnamefont {J.}~\bibnamefont
  {Planer}}, \bibinfo {author} {\bibfnamefont {J.}~\bibnamefont {Redinger}},
  \bibinfo {author} {\bibfnamefont {A.}~\bibnamefont {Gr\"uneis}}, \ and\
  \bibinfo {author} {\bibfnamefont {F.}~\bibnamefont {Mittendorfer}},\ }\href
  {\doibase 10.1103/PhysRevResearch.2.043361} {\bibfield  {journal} {\bibinfo
  {journal} {Phys. Rev. Research}\ }\textbf {\bibinfo {volume} {2}},\ \bibinfo
  {pages} {043361} (\bibinfo {year} {2020})}\BibitemShut {NoStop}%
\bibitem [{\citenamefont {Kaltak}\ \emph {et~al.}(2014)\citenamefont {Kaltak},
  \citenamefont {Klime\v{s}},\ and\ \citenamefont
  {Kresse}}]{PhysRevB.90.054115}%
  \BibitemOpen
  \bibfield  {author} {\bibinfo {author} {\bibfnamefont {M.}~\bibnamefont
  {Kaltak}}, \bibinfo {author} {\bibfnamefont {J.}~\bibnamefont {Klime\v{s}}},
  \ and\ \bibinfo {author} {\bibfnamefont {G.}~\bibnamefont {Kresse}},\ }\href
  {\doibase 10.1103/PhysRevB.90.054115} {\bibfield  {journal} {\bibinfo
  {journal} {Phys. Rev. B}\ }\textbf {\bibinfo {volume} {90}},\ \bibinfo
  {pages} {054115} (\bibinfo {year} {2014})}\BibitemShut {NoStop}%
\bibitem [{\citenamefont {Liu}\ \emph {et~al.}(2016)\citenamefont {Liu},
  \citenamefont {Kaltak}, \citenamefont {Klime\v{s}},\ and\ \citenamefont
  {Kresse}}]{PhysRevB.94.165109}%
  \BibitemOpen
  \bibfield  {author} {\bibinfo {author} {\bibfnamefont {P.}~\bibnamefont
  {Liu}}, \bibinfo {author} {\bibfnamefont {M.}~\bibnamefont {Kaltak}},
  \bibinfo {author} {\bibfnamefont {J.}~\bibnamefont {Klime\v{s}}}, \ and\
  \bibinfo {author} {\bibfnamefont {G.}~\bibnamefont {Kresse}},\ }\href
  {\doibase 10.1103/PhysRevB.94.165109} {\bibfield  {journal} {\bibinfo
  {journal} {Phys. Rev. B}\ }\textbf {\bibinfo {volume} {94}},\ \bibinfo
  {pages} {165109} (\bibinfo {year} {2016})}\BibitemShut {NoStop}%
\bibitem [{\citenamefont {Ramberger}\ \emph {et~al.}(2017)\citenamefont
  {Ramberger}, \citenamefont {Sch\"afer},\ and\ \citenamefont
  {Kresse}}]{PhysRevLett.118.106403}%
  \BibitemOpen
  \bibfield  {author} {\bibinfo {author} {\bibfnamefont {B.}~\bibnamefont
  {Ramberger}}, \bibinfo {author} {\bibfnamefont {T.}~\bibnamefont
  {Sch\"afer}}, \ and\ \bibinfo {author} {\bibfnamefont {G.}~\bibnamefont
  {Kresse}},\ }\href {\doibase 10.1103/PhysRevLett.118.106403} {\bibfield
  {journal} {\bibinfo  {journal} {Phys. Rev. Lett.}\ }\textbf {\bibinfo
  {volume} {118}},\ \bibinfo {pages} {106403} (\bibinfo {year}
  {2017})}\BibitemShut {NoStop}%
\bibitem [{\citenamefont {Jinnouchi}\ \emph
  {et~al.}(2019{\natexlab{a}})\citenamefont {Jinnouchi}, \citenamefont
  {Karsai},\ and\ \citenamefont {Kresse}}]{PhysRevB.100.014105}%
  \BibitemOpen
  \bibfield  {author} {\bibinfo {author} {\bibfnamefont {R.}~\bibnamefont
  {Jinnouchi}}, \bibinfo {author} {\bibfnamefont {F.}~\bibnamefont {Karsai}}, \
  and\ \bibinfo {author} {\bibfnamefont {G.}~\bibnamefont {Kresse}},\ }\href
  {\doibase 10.1103/PhysRevB.100.014105} {\bibfield  {journal} {\bibinfo
  {journal} {Phys. Rev. B}\ }\textbf {\bibinfo {volume} {100}},\ \bibinfo
  {pages} {014105} (\bibinfo {year} {2019}{\natexlab{a}})}\BibitemShut
  {NoStop}%
\bibitem [{\citenamefont {Bishop}(2006)}]{book_Bishop}%
  \BibitemOpen
  \bibfield  {author} {\bibinfo {author} {\bibfnamefont {C.~M.}\ \bibnamefont
  {Bishop}},\ }\href@noop {} {\emph {\bibinfo {title} {Pattern Recognition and
  Machine Learning (Information Science and Statistics)}}}\ (\bibinfo
  {publisher} {Springer},\ \bibinfo {address} {New York},\ \bibinfo {year}
  {2006})\BibitemShut {NoStop}%
\bibitem [{\citenamefont {Jinnouchi}\ \emph
  {et~al.}(2019{\natexlab{b}})\citenamefont {Jinnouchi}, \citenamefont
  {Lahnsteiner}, \citenamefont {Karsai}, \citenamefont {Kresse},\ and\
  \citenamefont {Bokdam}}]{PhysRevLett.122.225701}%
  \BibitemOpen
  \bibfield  {author} {\bibinfo {author} {\bibfnamefont {R.}~\bibnamefont
  {Jinnouchi}}, \bibinfo {author} {\bibfnamefont {J.}~\bibnamefont
  {Lahnsteiner}}, \bibinfo {author} {\bibfnamefont {F.}~\bibnamefont {Karsai}},
  \bibinfo {author} {\bibfnamefont {G.}~\bibnamefont {Kresse}}, \ and\ \bibinfo
  {author} {\bibfnamefont {M.}~\bibnamefont {Bokdam}},\ }\href {\doibase
  10.1103/PhysRevLett.122.225701} {\bibfield  {journal} {\bibinfo  {journal}
  {Phys. Rev. Lett.}\ }\textbf {\bibinfo {volume} {122}},\ \bibinfo {pages}
  {225701} (\bibinfo {year} {2019}{\natexlab{b}})}\BibitemShut {NoStop}%
\bibitem [{\citenamefont {Jinnouchi}\ \emph
  {et~al.}(2020{\natexlab{a}})\citenamefont {Jinnouchi}, \citenamefont
  {Karsai}, \citenamefont {Verdi}, \citenamefont {Asahi},\ and\ \citenamefont
  {Kresse}}]{doi:10.1063/5.0009491}%
  \BibitemOpen
  \bibfield  {author} {\bibinfo {author} {\bibfnamefont {R.}~\bibnamefont
  {Jinnouchi}}, \bibinfo {author} {\bibfnamefont {F.}~\bibnamefont {Karsai}},
  \bibinfo {author} {\bibfnamefont {C.}~\bibnamefont {Verdi}}, \bibinfo
  {author} {\bibfnamefont {R.}~\bibnamefont {Asahi}}, \ and\ \bibinfo {author}
  {\bibfnamefont {G.}~\bibnamefont {Kresse}},\ }\href {\doibase
  10.1063/5.0009491} {\bibfield  {journal} {\bibinfo  {journal} {J. Chem.
  Phys.}\ }\textbf {\bibinfo {volume} {152}},\ \bibinfo {pages} {234102}
  (\bibinfo {year} {2020}{\natexlab{a}})}\BibitemShut {NoStop}%
\bibitem [{\citenamefont {Jinnouchi}\ \emph
  {et~al.}(2020{\natexlab{b}})\citenamefont {Jinnouchi}, \citenamefont {Miwa},
  \citenamefont {Karsai}, \citenamefont {Kresse},\ and\ \citenamefont
  {Asahi}}]{doi:10.1021/acs.jpclett.0c01061}%
  \BibitemOpen
  \bibfield  {author} {\bibinfo {author} {\bibfnamefont {R.}~\bibnamefont
  {Jinnouchi}}, \bibinfo {author} {\bibfnamefont {K.}~\bibnamefont {Miwa}},
  \bibinfo {author} {\bibfnamefont {F.}~\bibnamefont {Karsai}}, \bibinfo
  {author} {\bibfnamefont {G.}~\bibnamefont {Kresse}}, \ and\ \bibinfo {author}
  {\bibfnamefont {R.}~\bibnamefont {Asahi}},\ }\href {\doibase
  10.1021/acs.jpclett.0c01061} {\bibfield  {journal} {\bibinfo  {journal} {The
  Journal of Physical Chemistry Letters}\ }\textbf {\bibinfo {volume} {11}},\
  \bibinfo {pages} {6946} (\bibinfo {year} {2020}{\natexlab{b}})}\BibitemShut
  {NoStop}%
\bibitem [{\citenamefont {Liu}\ \emph {et~al.}(2021)\citenamefont {Liu},
  \citenamefont {Verdi}, \citenamefont {Karsai},\ and\ \citenamefont
  {Kresse}}]{Peitao_2020}%
  \BibitemOpen
  \bibfield  {author} {\bibinfo {author} {\bibfnamefont {P.}~\bibnamefont
  {Liu}}, \bibinfo {author} {\bibfnamefont {C.}~\bibnamefont {Verdi}}, \bibinfo
  {author} {\bibfnamefont {F.}~\bibnamefont {Karsai}}, \ and\ \bibinfo {author}
  {\bibfnamefont {G.}~\bibnamefont {Kresse}},\ }\href {\doibase
  10.1103/PhysRevMaterials.5.053804} {\bibfield  {journal} {\bibinfo  {journal}
  {Phys. Rev. Materials}\ }\textbf {\bibinfo {volume} {5}},\ \bibinfo {pages}
  {053804} (\bibinfo {year} {2021})}\BibitemShut {NoStop}%
\bibitem [{\citenamefont {Verdi}\ \emph {et~al.}(2021)\citenamefont {Verdi},
  \citenamefont {Karsai}, \citenamefont {Liu}, \citenamefont {Jinnouchi},\ and\
  \citenamefont {Kresse}}]{Carla_2021}%
  \BibitemOpen
  \bibfield  {author} {\bibinfo {author} {\bibfnamefont {C.}~\bibnamefont
  {Verdi}}, \bibinfo {author} {\bibfnamefont {F.}~\bibnamefont {Karsai}},
  \bibinfo {author} {\bibfnamefont {P.}~\bibnamefont {Liu}}, \bibinfo {author}
  {\bibfnamefont {R.}~\bibnamefont {Jinnouchi}}, \ and\ \bibinfo {author}
  {\bibfnamefont {G.}~\bibnamefont {Kresse}},\ }\href@noop {} {\  (\bibinfo
  {year} {npj Comput. Mater., accepted, 2021})}\BibitemShut {NoStop}%
\bibitem [{\citenamefont {Allen}\ and\ \citenamefont
  {Tildesley}()}]{book_Allen_Tildesley}%
  \BibitemOpen
  \bibfield  {author} {\bibinfo {author} {\bibfnamefont {M.~P.}\ \bibnamefont
  {Allen}}\ and\ \bibinfo {author} {\bibfnamefont {D.~J.}\ \bibnamefont
  {Tildesley}},\ }\href@noop {} {\emph {\bibinfo {title} {Computer simulation
  of liquids (Oxford university press: New York, 1991)}}}\BibitemShut {NoStop}%
\bibitem [{\citenamefont {Parrinello}\ and\ \citenamefont
  {Rahman}(1980)}]{PhysRevLett.45.1196}%
  \BibitemOpen
  \bibfield  {author} {\bibinfo {author} {\bibfnamefont {M.}~\bibnamefont
  {Parrinello}}\ and\ \bibinfo {author} {\bibfnamefont {A.}~\bibnamefont
  {Rahman}},\ }\href {\doibase 10.1103/PhysRevLett.45.1196} {\bibfield
  {journal} {\bibinfo  {journal} {Phys. Rev. Lett.}\ }\textbf {\bibinfo
  {volume} {45}},\ \bibinfo {pages} {1196} (\bibinfo {year}
  {1980})}\BibitemShut {NoStop}%
\bibitem [{\citenamefont {Mahoney}\ and\ \citenamefont
  {Drineas}(2009)}]{Mahoney697}%
  \BibitemOpen
  \bibfield  {author} {\bibinfo {author} {\bibfnamefont {M.~W.}\ \bibnamefont
  {Mahoney}}\ and\ \bibinfo {author} {\bibfnamefont {P.}~\bibnamefont
  {Drineas}},\ }\href {\doibase 10.1073/pnas.0803205106} {\bibfield  {journal}
  {\bibinfo  {journal} {Proc. Natl. Acad. Sci. U.S.A.}\ }\textbf {\bibinfo
  {volume} {106}},\ \bibinfo {pages} {697} (\bibinfo {year}
  {2009})}\BibitemShut {NoStop}%
\bibitem [{\citenamefont {Stefanovich}\ \emph {et~al.}(1994)\citenamefont
  {Stefanovich}, \citenamefont {Shluger},\ and\ \citenamefont
  {Catlow}}]{PhysRevB.49.11560}%
  \BibitemOpen
  \bibfield  {author} {\bibinfo {author} {\bibfnamefont {E.~V.}\ \bibnamefont
  {Stefanovich}}, \bibinfo {author} {\bibfnamefont {A.~L.}\ \bibnamefont
  {Shluger}}, \ and\ \bibinfo {author} {\bibfnamefont {C.~R.~A.}\ \bibnamefont
  {Catlow}},\ }\href {\doibase 10.1103/PhysRevB.49.11560} {\bibfield  {journal}
  {\bibinfo  {journal} {Phys. Rev. B}\ }\textbf {\bibinfo {volume} {49}},\
  \bibinfo {pages} {11560} (\bibinfo {year} {1994})}\BibitemShut {NoStop}%
\bibitem [{\citenamefont {Aldebert}\ and\ \citenamefont
  {Traverse}(1985)}]{https://doi.org/10.1111/j.1151-2916.1985.tb15247.x}%
  \BibitemOpen
  \bibfield  {author} {\bibinfo {author} {\bibfnamefont {P.}~\bibnamefont
  {Aldebert}}\ and\ \bibinfo {author} {\bibfnamefont {J.-P.}\ \bibnamefont
  {Traverse}},\ }\href {\doibase
  https://doi.org/10.1111/j.1151-2916.1985.tb15247.x} {\bibfield  {journal}
  {\bibinfo  {journal} {J. Am. Ceram. Soc.}\ }\textbf {\bibinfo {volume}
  {68}},\ \bibinfo {pages} {34} (\bibinfo {year} {1985})}\BibitemShut {NoStop}%
\bibitem [{\citenamefont {Bart\'{o}k}\ and\ \citenamefont
  {Yates}(2019)}]{doi:10.1063/1.5094646}%
  \BibitemOpen
  \bibfield  {author} {\bibinfo {author} {\bibfnamefont {A.~P.}\ \bibnamefont
  {Bart\'{o}k}}\ and\ \bibinfo {author} {\bibfnamefont {J.~R.}\ \bibnamefont
  {Yates}},\ }\href {\doibase 10.1063/1.5094646} {\bibfield  {journal}
  {\bibinfo  {journal} {J. Chem. Phys.}\ }\textbf {\bibinfo {volume} {150}},\
  \bibinfo {pages} {161101} (\bibinfo {year} {2019})}\BibitemShut {NoStop}%
\bibitem [{\citenamefont {Furness}\ \emph {et~al.}(2020)\citenamefont
  {Furness}, \citenamefont {Kaplan}, \citenamefont {Ning}, \citenamefont
  {Perdew},\ and\ \citenamefont {Sun}}]{doi:10.1021/acs.jpclett.0c02405}%
  \BibitemOpen
  \bibfield  {author} {\bibinfo {author} {\bibfnamefont {J.~W.}\ \bibnamefont
  {Furness}}, \bibinfo {author} {\bibfnamefont {A.~D.}\ \bibnamefont {Kaplan}},
  \bibinfo {author} {\bibfnamefont {J.}~\bibnamefont {Ning}}, \bibinfo {author}
  {\bibfnamefont {J.~P.}\ \bibnamefont {Perdew}}, \ and\ \bibinfo {author}
  {\bibfnamefont {J.}~\bibnamefont {Sun}},\ }\href {\doibase
  10.1021/acs.jpclett.0c02405} {\bibfield  {journal} {\bibinfo  {journal} {The
  Journal of Physical Chemistry Letters}\ }\textbf {\bibinfo {volume} {11}},\
  \bibinfo {pages} {8208} (\bibinfo {year} {2020})}\BibitemShut {NoStop}%
\bibitem [{\citenamefont {Momma}\ and\ \citenamefont
  {Izumi}(2011)}]{Momma:db5098}%
  \BibitemOpen
  \bibfield  {author} {\bibinfo {author} {\bibfnamefont {K.}~\bibnamefont
  {Momma}}\ and\ \bibinfo {author} {\bibfnamefont {F.}~\bibnamefont {Izumi}},\
  }\href {\doibase 10.1107/S0021889811038970} {\bibfield  {journal} {\bibinfo
  {journal} {Journal of Applied Crystallography}\ }\textbf {\bibinfo {volume}
  {44}},\ \bibinfo {pages} {1272} (\bibinfo {year} {2011})}\BibitemShut
  {NoStop}%
\end{thebibliography}%

\end{document}


\title{Supplementary Material to ``Phase Transitions of Zirconia: Machine-Learned Force Fields Beyond Density Functional Theory"}

\author{Peitao Liu}
\email{peitao.liu@univie.ac.at}
\affiliation{VASP Software GmbH, Sensengasse 8, 1090 Vienna, Austria}

\author{Carla Verdi}
\affiliation{University of Vienna, Faculty of Physics and Center for
Computational Materials Science, Kolingasse 14-16, A-1090, Vienna,
Austria}

\author{Ferenc Karsai}
\affiliation{VASP Software GmbH, Sensengasse 8, 1090 Vienna, Austria}

\author{Georg Kresse}
\affiliation{University of Vienna, Faculty of Physics and Center for
Computational Materials Science, Kolingasse 14-16, A-1090, Vienna,
Austria}
\affiliation{VASP Software GmbH, Sensengasse 8, 1090 Vienna, Austria}

\maketitle

\section{First-principles calculations}\label{sec:DFT_details}

All first-principles calculations were performed using the projector augmented wave (PAW) method~\cite{PhysRevB.50.17953}
as implemented in the Vienna \emph{Ab initio} Simulation Package (VASP)~\cite{PhysRevB.47.558, PhysRevB.54.11169}.
For density functional theory (DFT) calculations, the standard PAW potentials ({\tt Zr\_sv} and {\tt O}) and a plane wave cutoff of 600 eV were employed.
The electronic interactions were described using the Perdew-Burke-Ernzerhof functional (PBE)~\cite{PhysRevLett.77.3865}
and the strongly constrained appropriately normed functional (SCAN)~\cite{PhysRevLett.115.036402}.
A $\Gamma$-centered $k$-point grid with a spacing of 0.31 $\AA^{-1}$ between $k$ points
(corresponding to a 2$\times$2$\times$2 $k$-point grid for a 96-atom cell) was employed.
The Gaussian smearing method with a smearing width of 0.05 eV was used.
Whenever ground state structures were required, the electronic optimization was performed until the total energy difference between two iterations was less than 10$^{-6}$ eV.
The structures were optimized until the forces were smaller than 5 meV/\AA.
The phonon dispersions were calculated by finite displacements using the Phonopy code~\cite{PhysRevB.78.134106}.
For all phonon calculations, a 96-atom supercell and a $2\times2\times2$ $k$-point grid were used.
To account for the long-range dipole-dipole force constants, the nonanalytic contribution
to the dynamical matrix was treated using the method of Ref.~\cite{PhysRevB.55.10355}.
The static dielectric tensor and atomic Born effective charges were calculated using the PBEsol functional~\cite{PhysRevLett.100.136406}.

\begin{figure}
\begin{center}
\includegraphics[width=0.47\textwidth, trim = {0 0.05cm 0.2cm  0.2cm}, clip]{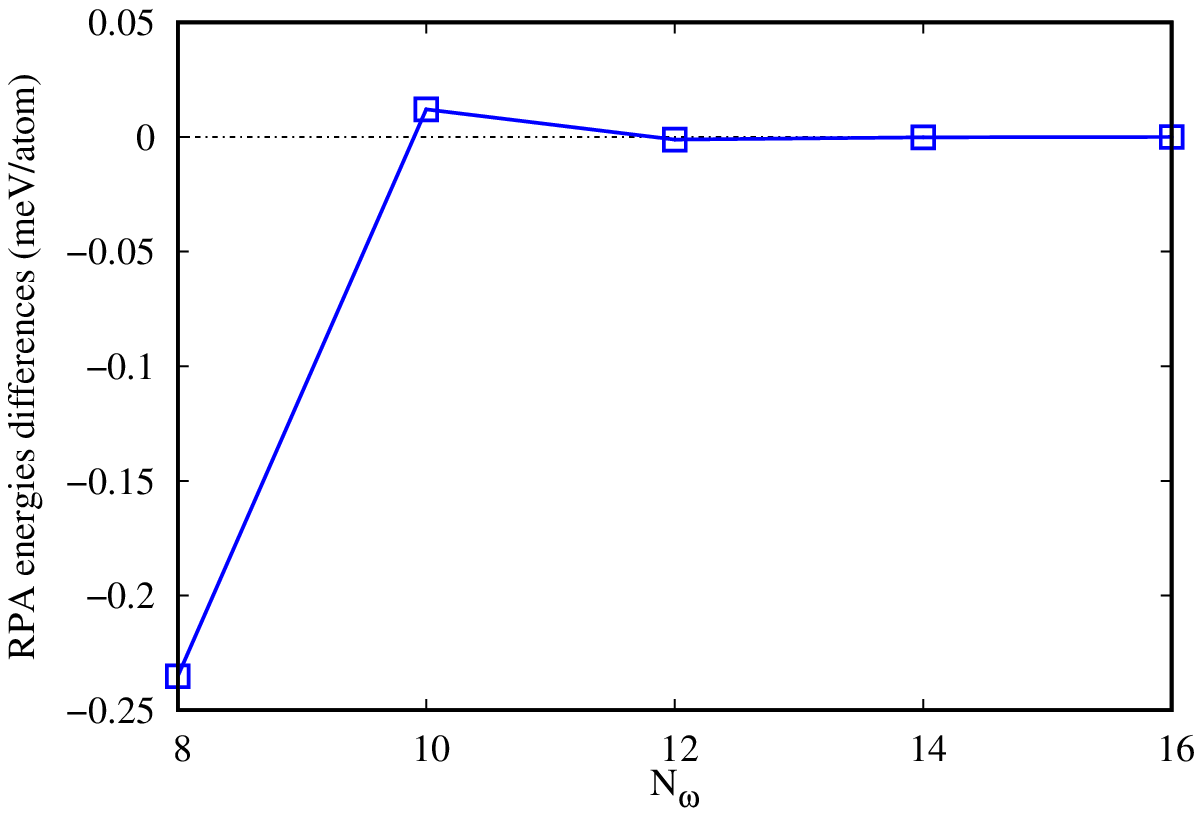}
\end{center}
\caption{Calculated RPA energies (in meV/atom) of a monoclinic structure of ZrO$_2$ with 24 atoms in the cell
as a function of the employed number of imaginary/frequency points $N_{\omega}$.
Note that the RPA energy obtained using $N_{\omega}$=16 is taken as reference.
The RPA energy obtained using $N_{\omega}=10$ converges to within 0.012 meV/atom.
 }
\label{RPA_energy_wrt_NOMEGA16}
\end{figure}

\begin{figure}
\begin{center}
\includegraphics[width=0.47\textwidth, trim = {0 0.05cm 0.2cm  0.2cm}, clip]{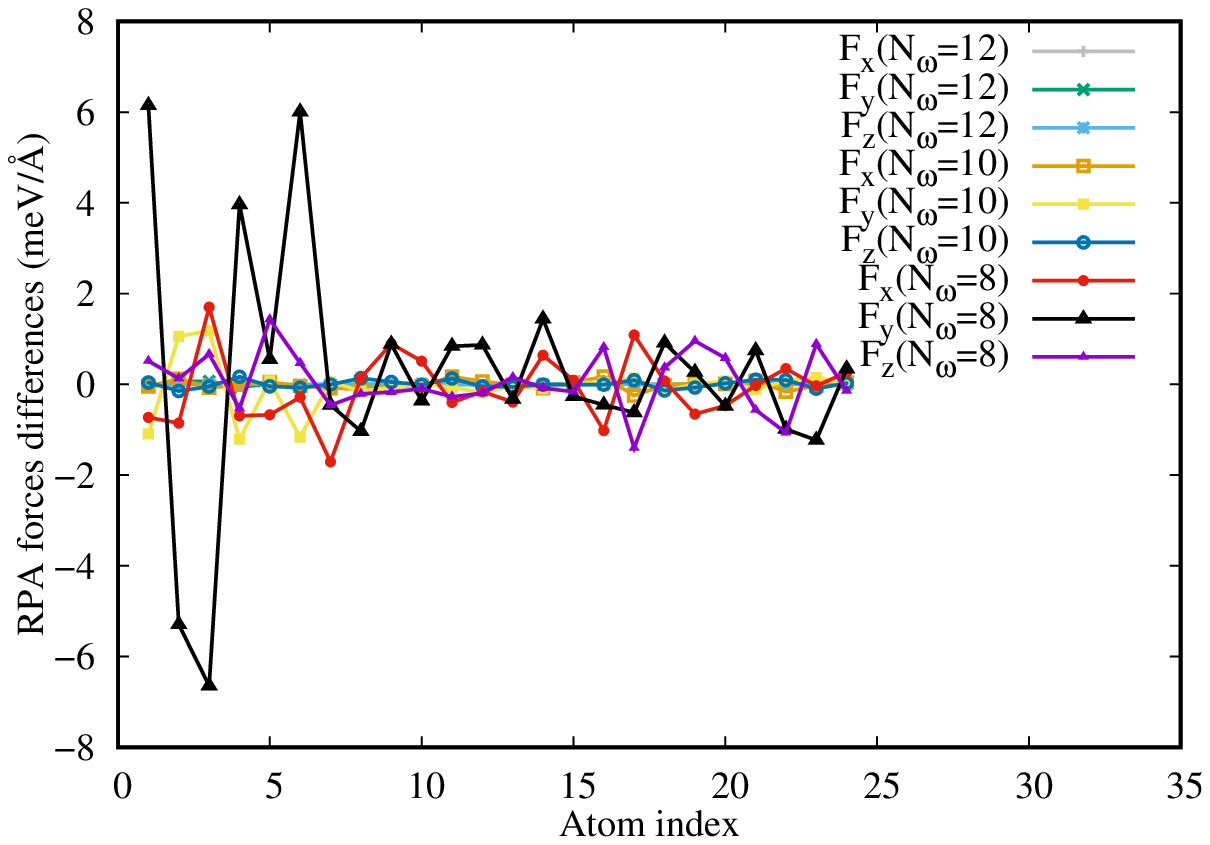}
\end{center}
\caption{Calculated RPA forces (in meV/\AA) of each atom in a monoclinic structure of ZrO$_2$ with 24 atoms in the cell
referenced to the ones obtained using $N_{\omega}=16$.
The results achieved by three $N_{\omega}$ (12, 10, and 8) are shown. One can observe that the forces obtained using $N_{\omega}=10$
are converged to within 1 meV/\AA.
 }
\label{RPA_force_wrt_NOMEGA16}
\end{figure}

\begin{figure}
\begin{center}
\includegraphics[width=0.47\textwidth, trim = {0 0.05cm 0.2cm  0.2cm}, clip]{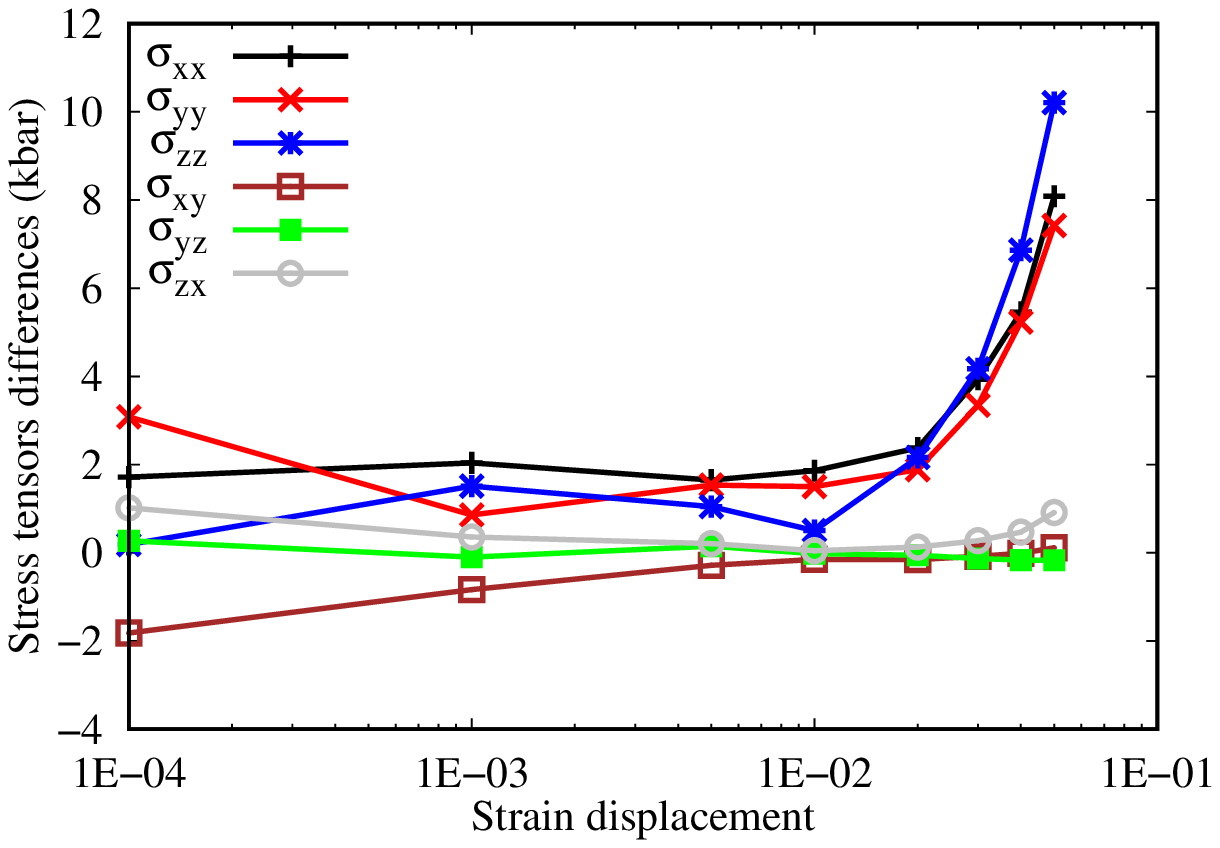}
\end{center}
\caption{Stress tensors (in kbar) calculated by finite differences at a plane wave cutoff of 520 eV
 referenced to the ones calculated internally by VASP using a plane wave cutoff of 1600 eV
[($\sigma_{xx}$, $\sigma_{yy}$, $\sigma_{zz}$, $\sigma_{xy}$, $\sigma_{yz}$, $\sigma_{zx}$)=(79.15,  132.58,  58.07,  $-$5.52,  $ -$22.18,  49.09) kbar]
as a function of strain displacement. The very large plane wave cutoff of 1600 eV was chosen to minimize the Pulay stress.
A finite temperature MD snapshot of m-ZrO$_2$ with 24 atoms in the cell was employed for this test.
A similar behavior was observed for the tetragonal as well as cubic phase of ZrO$_2$ (not shown). }
\label{stress_vs_displacement}
\end{figure}

For the random phase approximation (RPA) calculations, the GW PAW potentials ({\tt Zr\_sv\_GW} and {\tt O\_GW})
were used. These GW PAW potentials are constructed by using additional projectors above the vacuum level,
and therefore, they describe well the high-energy scattering properties of the atoms and are more accurate for the polarizability-dependent RPA calculations.
As shown in Ref.~\cite{PhysRevResearch.2.043361}, the RPA is capable to simultaneously describe well both the structural and electronic properties of ZrO$_2$.
Due to the large computational cost involved in RPA calculations, a reduced plane wave cutoff of 520 eV and
a `bcc'-like generalized regular grid with 8 $k$ points in the full Brillouin zone
were used. The RPA energies and forces are calculated using an efficient low-scaling algorithm~\cite{PhysRevB.90.054115,PhysRevB.94.165109,PhysRevLett.118.106403}.
The energy cutoff for the response function was chosen to be the same as  the plane wave cutoff (520 eV)
and the number of imaginary time/frequency points ($N_{\omega}$) was set to 10,
which is sufficient to ensure convergence of RPA energies (see Fig.~\ref{RPA_energy_wrt_NOMEGA16}) and forces (see Fig.~\ref{RPA_force_wrt_NOMEGA16}).
The stress tensor $\sigma_{\alpha\beta}$ at the RPA level was calculated  via finite differences of the RPA total energies using twelve slightly distorted structures through
\begin{equation}\label{eq:stress_tensor}
\sigma_{\alpha\beta}\approx
 -\frac{1}{\Omega} \frac{E(e_{\alpha\beta}=+\delta)-E(e_{\alpha\beta}=-\delta)}{2\delta},
\end{equation}
where $\alpha$ and $\beta$ represent the Cartesian coordinate indices, $\Omega$ is the system volume, and $e_{\alpha\beta}$ is the strain tensor.
Here, $\delta=0.02$ was adopted. This value on the one hand ensures the convergence of the stress tensors for PBE calculations
to within 2.5 kbar (see Fig.~\ref{stress_vs_displacement}), and
on the other hand is sufficiently large to minimize the shell effects (plane wave $\lG$-vectors moving in and out of the cutoff sphere) for RPA calculations. We note that the stress tensors calculated by the finite difference method is less prone to the Pulay stress than those calculated internally within VASP. The latter often suffers from basis set incompleteness errors because of the fixed plane wave basis set when the cell is distorted. This is the reason why we chose to use a relatively large plane wave cutoff of 600 eV for the DFT calculations, whose stress tensors are directly calculated by using the VASP internal routines. Even then the diagonal components of the stress tensor are corrected by the calculated Pulay stress before generating the machine-learned force fields (MLFFs). The phonon frequencies at $\Gamma$ predicted by RPA were calculated by finite differences, with
the long-range dipole-dipole interactions calculated at the level of PBEsol.

\section{MLFF training}\label{sec:training_details}

Our MLFFs were initially trained on-the-fly during FP molecular dynamics (MD) simulations
based on the Bayesian linear regression~\cite{PhysRevB.100.014105,book_Bishop}.
For a comprehensive description of the on-the-fly MLFF generation implemented in VASP, we refer to Refs.~\cite{PhysRevLett.122.225701,PhysRevB.100.014105,doi:10.1063/5.0009491}.
A concise summary of this method can be found in Refs.~\cite{doi:10.1021/acs.jpclett.0c01061,Peitao_2020}.
For the on-the-fly training, the PBEsol functional~\cite{PhysRevLett.100.136406} was used,
since it predicts accurate lattice parameters for all the three phases of ZrO$_2$~\cite{Carla_2021} on par with SCAN, but it is cheaper.

The FPMD simulations were performed in the isothermal-isobaric (NPT) ensemble at ambient pressure using a Langevin thermostat~\cite{book_Allen_Tildesley}
combined with the Parrinello-Raman method~\cite{PhysRevLett.45.1196}. The time step was set to 2.5 fs.
For the MLFF generation, the separable descriptors~\cite{doi:10.1063/5.0009491} were used.
The cutoff radius for the three-body descriptors and the width of the Gaussian functions used for broadening the atomic
distributions of the three-body descriptors were set to 6 $\AA$ and 0.4 $\AA$, respectively. The number of radial basis
functions and maximum three-body momentum quantum number of the spherical harmonics used to expand the atomic distribution
for the three-body descriptors were set to 15 and 4, respectively. The parameters for the two-body descriptors were
the same as those for the three-body descriptor.

As mentioned in the main text, two MLFFs were constructed on the fly.
The first MLFF was trained on a 96-atom cell, whereas the second one was trained on a smaller 24-atom cell.
We note that the second MLFF was not used and this training step was just used to collect a new reference dataset,
from which a subset of structures were extracted using a singular value decomposition (SVD) rank compression of the kernel matrix using a CUR algorithm~\cite{Mahoney697,PhysRevB.100.014105}
and then recalculated by high-level QM calculations.
For the first MLFF, the detailed training procedures can be found in Ref.~\cite{Carla_2021}, and in the end, 592 structures were collected in the training dataset.
For the second MLFF trained on smaller unit cells, the following training strategy was employed.
($i$) We first trained the force field by heating the monoclinic ZrO$_2$ from 0 K to 1800 K using 20 000 MD steps starting from the DFT relaxed structure.
($ii$) Then, we continued training the tetragonal phase by a heating run from 1700 K to 2600 K using 10 000 MD steps.
($iii$) The force field was further trained  by heating the cubic phase from 2500 K to 2800 K using 3333 MD steps.
Note that in steps ($ii)$ and ($iii$), the initial structures were obtained by equilibrating the tetragonal and cubic structures
at 1700 K and 2500 K, respectively, using the on-the-fly MLFF scheme, but the thus generated MLFFs were discarded.
($iv$) To include the ideal tetragonal and cubic structures at 0 K, additional short heating runs from 0 K to 10 K using 100 MD steps
were performed starting from the DFT relaxed  tetragonal and cubic structures, respectively.
Eventually, only 1275 FP calculations were performed out of 33 533 MD steps, i.e., nearly 96.2\%  of the FP calculations were bypassed.

\begin{table*}
  \caption{ Zero-temperature structural parameters of all three phases of ZrO$_2$ and energy differences between
  phases predicted by different exchange-correlation functionals and MLFFs. $d_z$ represents the displacement of the oxygen (in unit of $c$) atoms along the $z$ direction with respect
 to ideal cubic position, and $\beta$ is the angle between the lattice vectors $\bf a$ and   $\bf c$ in the monoclinic phase.
 The values in parentheses are RPA predicted energy differences for the structures that are optimized by MLFF-RPA$^{\Delta}$.
The experimental structural parameters and volumes are extrapolated to 0 K~\cite{PhysRevB.49.11560,https://doi.org/10.1111/j.1151-2916.1985.tb15247.x}.}
  \begin{ruledtabular}
 \begin{tabular}{lccccccccc}
             & PBE &   MLFF-PBE & SCAN & MLFF-SCAN & MLFF-SCAN$^{\Delta}$ & MLFF-RPA$^{\Delta}$& Expt. \\
Monoclinic                            &       &           &        &       &        &         &       \\
$a$ (\AA)                             & 5.194 &  5.189    &  5.150 & 5.152 & 5.155  &  5.146  & 5.151 \\
$b$ (\AA)                             & 5.248 &  5.251    &  5.225 & 5.223 & 5.220  &  5.219  & 5.212 \\
$c$ (\AA)                             & 5.381 &  5.378    &  5.326 & 5.329 & 5.335  &  5.313  & 5.317 \\
$\beta$ (deg.)                        & 99.66 &  99.68    &  99.35 & 99.40 & 99.40  &  99.39  & 99.23 \\
Volume ($\AA^3$/f.u.)                 & 36.15 &  36.11    &  35.35 & 35.37 & 35.41  &  35.20  & 35.22 \\
\\
Tetragonal                            &       &           &        &       &        &         &       \\
$a$ (\AA)                             & 3.624 &  3.625    &  3.600 & 3.600 & 3.601  &  3.592  & 3.571 \\
$c$ (\AA)                             & 5.284 &  5.290    &  5.220 & 5.230 & 5.231  &  5.189  & 5.182 \\
$c/a$                                 & 1.458 &  1.459    &  1.450 & 1.453 & 1.453  &  1.445  & 1.451 \\
$d_z$                                 & 0.057 &  0.058    &  0.052 & 0.054 & 0.054  &  0.048  & 0.047 \\
Volume ($\AA^3$/f.u.)                 & 34.70 &  34.76    &  33.82 & 33.90 & 33.91  &  33.47  & 33.01 \\
$\Delta E_{t-m}$ (eV/f.u.) & 0.110 &  0.109    &  0.074 & 0.074 & 0.075  &  0.067 (0.069)  & --- \\
\\
Cubic                                 &       &           &        &       &        &         &       \\
$a$(\AA)                              & 5.120 &  5.126    &  5.088 & 5.090 & 5.091  &  5.076  & --- \\
Volume ($\AA^3$/f.u.)                 & 33.56 &  33.68    &  32.92 & 32.97 & 32.99  &  32.70  & --- \\
$\Delta E_{c-t}$ (eV/f.u.) & 0.102 &  0.100    &  0.085 & 0.083 & 0.083  &  0.053 (0.047)  & --- \\
 \end{tabular}
 \end{ruledtabular}
 \label{tab:lattice_constants}
\end{table*}

At the end of the on-the-fly training, we determined the final regression
coefficients
by using a factor of 10 for the relative weight of the energy equations with respect to the equations for the forces and stress tensors,
and a SVD to solve the least-squares problem.
These were found to improve the overall accuracy of MLFFs~\cite{Peitao_2020,Carla_2021}.

Finally, all the structures contained in $T^{(1)}$ and $T^{(3)}$ were recalculated by PBE and SCAN,
whereas RPA calculations were performed only for the structures in $T^{(3)}$ (including 168 structures of 24 atoms, see the main text).
MLFF-$\Delta$ was obtained by machine learning the differences in energies, forces and stress tensors between high-level and low-level QM calculations
using the separable descriptors~\cite{doi:10.1063/5.0009491} with low spatial resolution (0.8 $\AA$) and a small number of radial basis functions (8) for
both the radial and angular parts.

The MLFFs directly trained using PBE and SCAN are referred to as MLFF-PBE and MLFF-SCAN, respectively.
The MLFFs indirectly trained using SCAN and RPA via the $\Delta$-machine learning ($\Delta$-ML) approach are denoted as MLFF-SCAN$^\Delta$
and MLFF-RPA$^\Delta$, respectively.
We note that although it is possible to generate the MLFF-RPA$^\Delta$  based on the MLFF-PBE (constructed using a plane wave cutoff of 600 eV and standard PAW potentials), the RPA calculations in particular for the forces are rather demanding using such a large cutoff energy for structures of 24 atoms.
Therefore, it is expedient to generate a second MLFF-PBE using a reduced plane wave cutoff of 520 eV and the GW PAW potentials as in the RPA calculations,
on which the MLFF-RPA$^\Delta$ is built. We note that the changes in the cutoff energy (from 600 eV to 520 eV) and the potentials (from PAW to GW PAW) for ZrO$_2$
have negligible effects on the accuracy of the resulting MLFF-PBE if the stress tensors calculated by the VASP internal routines were corrected for the Pulay stress.
For generating MLFF-$\Delta$ where $\Delta=$RPA$-$PBE, both PBE and RPA calculations were performed on the structures in $T^{(3)}$
using the plane wave cutoff of 520 eV and the GW PAW potentials.

\section{MLFF validation}\label{sec:validation_details}

The MLFFs including MLFF-PBE, MLFF-SCAN, and MLFF-SCAN$^\Delta$
have been validated on a test dataset containing 120 structures of 96 atoms
(40 monoclinic structures at $T$=1000 K, 40 tetragonal structures at $T$=2000 K,  and 40 cubic structures at $T$=3000 K).
However, for MLFF-RPA$^\Delta$,
due to the high computational cost for the RPA calculations, a reduced test dataset containing 60 structures of 24 atoms (20 monoclinic structures
at $T$=1000 K, 20 tetragonal structures at $T$=2000 K,  and 20 cubic structures at $T$=3000 K) was used.
All the structures in the test datasets were generated from MD simulations using the NPT ensemble and MLFF-SCAN.

\begin{figure*}
\begin{center}
\includegraphics[width=0.85\textwidth, clip]{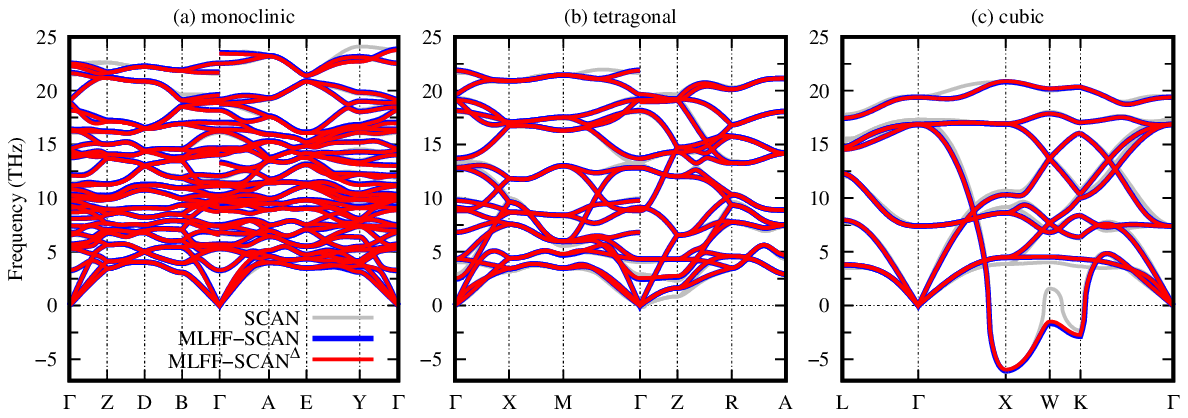}
\end{center}
\caption{Phonon dispersion relations of (a) monoclinic, (b) tetragonal, and (c) cubic  ZrO$_2$  at 0 K predicted by SCAN (grey lines), MLFF-SCAN (blue lines),
and MLFF-SCAN$^{\Delta}$ (red lines). Almost no difference is observed between MLFF-SCAN$^{\Delta}$ and MLFF-SCAN for all three phases,
indicating their comparable accuracies.
}
\label{fig:phonon_MLSCAN_SCAN}
\end{figure*}

\begin{figure*}
\begin{center}
\includegraphics[width=0.85\textwidth, clip]{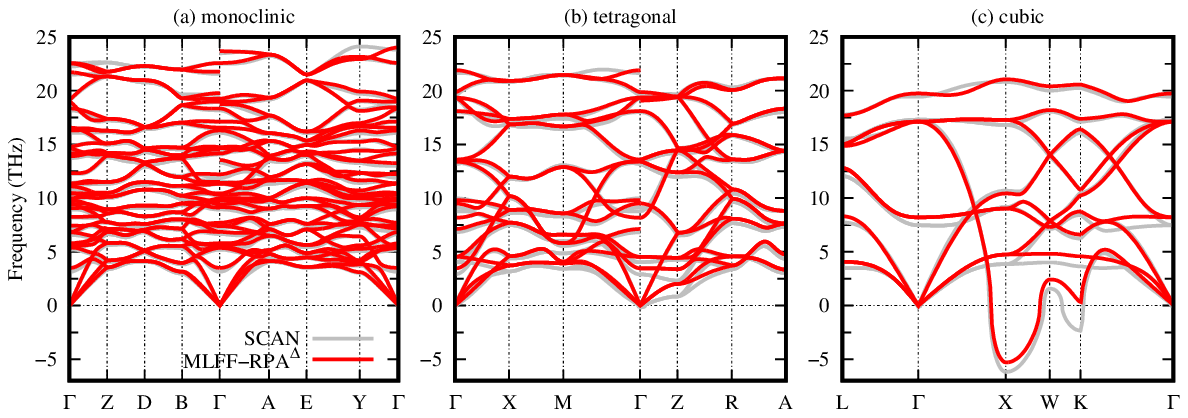}
\end{center}
\caption{Phonon dispersion relations of (a) monoclinic, (b) tetragonal, and (c) cubic  ZrO$_2$  at 0 K predicted by MLFF-RPA$^\Delta$ (red lines).
The results from SCAN (grey lines) are also given for comparison, showing very similar phonon dispersions with MLFF-RPA$^\Delta$ for all three phases.
}
\label{fig:phonon_MLRPA_SCAN}
\end{figure*}

As shown in Table I of the main text, all generated MLFFs are very accurate with small validation errors.
The errors for MLFF-SCAN are slightly larger than those for MLFF-PBE.
This is likely due to the poor numerical performance of the SCAN functional~\cite{doi:10.1063/1.5094646,doi:10.1021/acs.jpclett.0c02405}.
The MLFF-SCAN$^\Delta$ derived by the $\Delta$-ML approach
exhibits almost comparable accuracy as MLFF-SCAN that was directly trained by SCAN.
The good accuracies of the obtained MLFF-PBE and MLFF-SCAN
 have also been showcased in their good predictions of structural parameters,
energy differences between different phases (Table~\ref{tab:lattice_constants}), and phonon dispersion relations
(Fig.~\ref{fig:phonon_MLSCAN_SCAN}) as compared to their respective DFT counterparts.
As expected, PBE overestimates the lattice constants and the energy differences between the phases.
SCAN improves upon PBE because of the improved treatment of intermediate Van der Waals interactions.
Our DFT results are overall in good agreement with Ref.~\cite{PhysRevResearch.2.043361}.
From Table~\ref{tab:lattice_constants} one can also observe that MLFF-SCAN$^\Delta$
performs almost equally well as MLFF-SCAN in the prediction of structural parameters and energy differences between the phases,
consistent with the error analysis shown in Table I of the main text..
This is also true in predicting phonon dispersion relations, as demonstrated in Fig.~\ref{fig:phonon_MLSCAN_SCAN}.
Almost no difference is observed between MLFF-SCAN$^{\Delta}$ and MLFF-SCAN for all three phases,
validating the feasibility of the $\Delta$-ML approach.

\begin{figure*}
\begin{center}
\includegraphics[width=0.8\textwidth, clip]{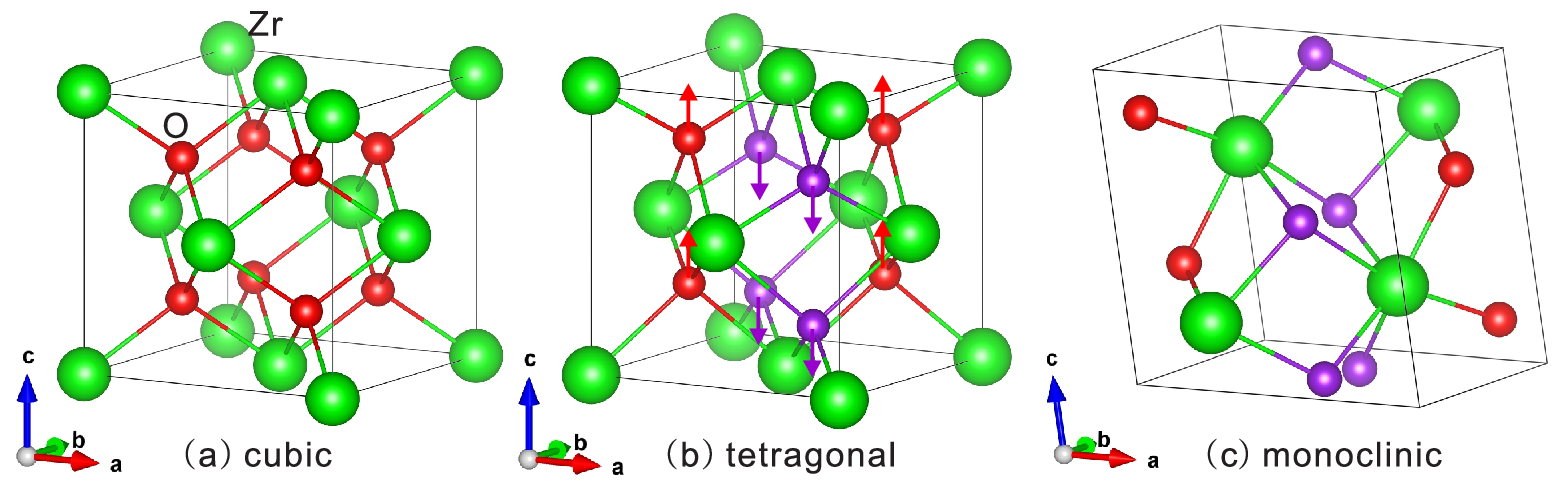}
\end{center}
\caption{Crystal structures of (a) cubic, (b) tetragonal, and (c) monoclinic ZrO$_2$. Large and small spheres denote Zr and O atoms, respectively.
The red and purple colors in the tetragonal structure (b) are used to distinguish O atoms that are displaced
upwards (red arrows) and downwards (purple arrows) as compared to the cubic structure, and in the monoclinic structure (c) they indicate the two nonequivalent O atoms.
 Structural models were generated using VESTA~\cite{Momma:db5098}.
 }
\label{fig:Str_plot_ZrO2}
\end{figure*}

The MLFF-RPA$^\Delta$ is validated on a reduced test dataset of small unit cells, i.e., 60 structures of 24 atoms,
because of the large computational cost of the RPA calculations. The validation errors in energies, forces and stress
tensors calculated by MLFF-RPA$^\Delta$ are given in Table II of the main text. For comparison, the validation errors
for MLFF-PBE and MLFF-SCAN are also shown.
First, one can observe that MLFF-RPA$^\Delta$  shows comparable validation errors as MLFF-PBE and MLFF-SCAN,
implying their comparably good accuracy. The slightly larger validation errors calculated by MLFF-RPA$^\Delta$
arise from the relatively noisy nature of RPA. Second, one may notice that the validation errors of energies and
stress tensors on small unit cells of 24 atoms for all the MLFFs (not just limited to MLFF-RPA ) are larger than those
on larger unit cells of 96 atoms, whereas the RMSEs for the forces remain almost  unchanged (compare Table II and Table I in the main text).
This difference in fact originates from the definition of RMSE and can be understood from the error propagation with respect to the system size.
Specifically, according to basic statistics, the variance of the difference between the DFT and MLFF
predicted energy Var$(E^{\rm DFT}-E^{\rm MLFF})$ is expected to be proportional to the system size $N$,
i.e., Var$(E^{\rm DFT}-E^{\rm MLFF})\sim N$, if the errors in the predicted local energies are statistically independent.
The RMSE of the energy per atom (in meV/atom) are calculated as
\begin{equation}\label{eq:RMSE_E}
{\rm RMSE}(E)= \sqrt{\frac{1}{M} \sum_i^M  \Big[(E^{\rm DFT}_i - E^{\rm MLFF}_i)/N_i\Big]^2},
\end{equation}
where $M$ is the number of structures and  $N_i$ is the number of atoms in the structure $i$.
From this definition, one can readily show that
\begin{equation}\label{eq:RMSE_E2}
{\rm RMSE}(E)= \sqrt{\frac{{\rm Var}(E^{\rm DFT}-E^{\rm MLFF})}{N^2}} \sim \frac{1}{\sqrt{N}},
\end{equation}
This means that if the system becomes $N$ times larger, then the RMSE of the energy per atom becomes  $1/\sqrt{N}$ times smaller.
This explains why all the MLFFs predict a larger RMSE for the energy per atom for the 24-atom cells than for the 96-atom cells.
The same error analysis holds for the stress tensor, since by definition it is calculated as the derivative
of the energy with respect to the strain and then divided by the system volume.
Nevertheless, this error propagation rule does not apply for the RMSE for the forces,
since the force is an intensive property that is independent of system size.

\section{Phase transitions of zirconia within the quasi-harmonic approximation}\label{sec:phase_transition_details}

\begin{figure}
\begin{center}
\includegraphics[width=0.49\textwidth, clip]{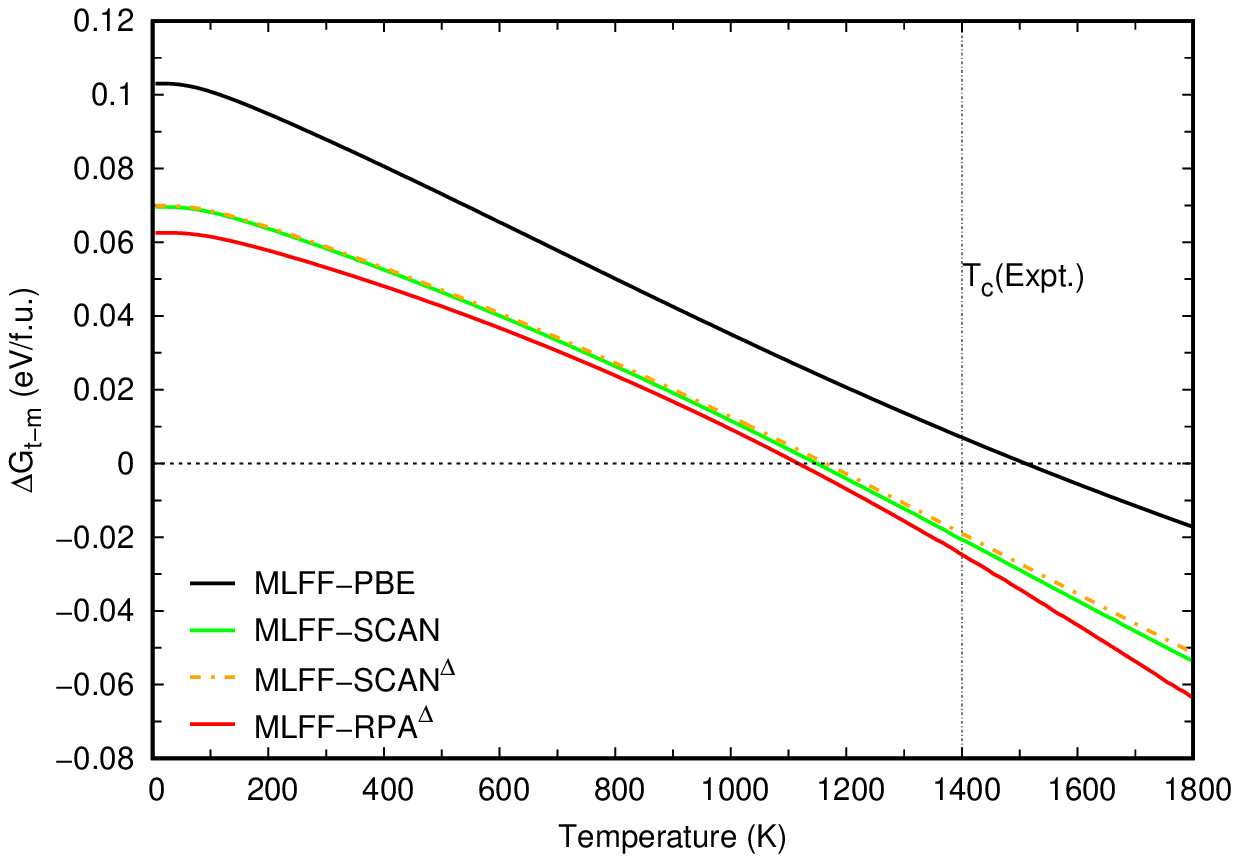}
\end{center}
\caption{The free energy difference (per formula unit) between the tetragonal and
monoclinic phase as a function of temperature predicted by the MLFFs within the quasi-harmonic approximation (QHA)
using the quantum Bose-Einstein (BE) statistics. We note that the results obtained from the BE statistics
and classical Maxwell-Boltzmann statistics are almost identical for temperatures above 200 K (not shown),
indicating that the quantum effect can be neglected at high temperatures.
 }
\label{fig:QHA}
\end{figure}

The primitive cells of the cubic, tetragonal, and monoclinic structures of
zirconia contain one, two, and four ZrO$_2$ formula units respectively.
A unit cell of 12 atoms can thus be used to accommodate all three phases (see Fig.~\ref{fig:Str_plot_ZrO2}).

As shown in the phonon dispersions (Fig.~\ref{fig:phonon_MLSCAN_SCAN}),
both monoclinic and tetragonal phases are dynamically stable at 0 K. In addition, no soft phonons are observed for both phases
when the volume is changed within 5\%. This allows us to estimate the $T_c$ using the quasi-harmonic approximation (QHA).
Fig.~\ref{fig:QHA} shows the free energy difference between the tetragonal and monoclinic phase
as a function of temperature predicted by the MLFFs within the QHA using the quantum Bose-Einstein statistics.
Accordingly, MLFF-PBE predicts a value of 1511 K for $T_c$.
MLFF-SCAN$^\Delta$ and MLFF-SCAN yield close values of 1148 K and 1164 K, respectively,
which are very close to the one obtained by the MLFF-PBEsol (1178 K) in Ref.~\cite{Carla_2021}.
In addition, for nearly the entire temperature range, MLFF-SCAN$^\Delta$  predicts very close free energies as compared to MLFF-SCAN.
This is expected, since both MLFF-SCAN$^\Delta$  and MLFF-SCAN overall show similar validation errors
as well as lattice parameters, energy differences between the phases, and phonon dispersions.
As compared to MLFF-SCAN, MLFF-RPA$^\Delta$  predicts a slightly smaller value of $T_c$ (about 1117 K).
In general, we find that within the QHA the predicted $T_c$ is correlated to the calculated energy differences between the two phases at 0 K (see Table~\ref{tab:lattice_constants}).

\section{SVD rank compression}\label{sec:SVD_CUR}

In order to select few most representative structures for machine-learning the differences from a large pool of dataset,
we employed SVD rank compression of the kernel matrix based on the leverage-score CUR algorithm~\cite{Mahoney697,PhysRevB.100.014105}.
In the following, the CUR algorithm is briefly introduced.

We denote $\mathbf{K}$ as a kernel matrix calculated in the feature space. It is a squared matrix whose elements $K_{ij}$ measure the similarity between two local reference configurations $i$ and $j$. The formulation of the CUR algorithm starts from the diagonalization of the matrix $\mathbf{K}$:
\begin{align}
\mathbf{U}^{\textit{T}}\mathbf{K}\mathbf{U}&=\mathbf{L}=\mathrm{diag}\left(l_{\mathrm{1}},...,l_{N}\right), \label{equation_diag}
\end{align}
where $N$ is the dimension of the matrix $\mathbf{K}$ and $\mathbf{U}$ is the eigenvector matrix defined as
\begin{align}
\mathbf{U}&=\left(\mathbf{u}_{\mathrm{1}},...,\mathbf{u}_{N}\right), \label{equation_U}\\
\mathbf{u}_{j}&=\left(u_{1j},...,u_{Nj}\right)^{\textit{T}}, \label{equation_u}
\end{align}
Eq.~(\ref{equation_diag}) can be rewritten as
\begin{align}
\mathbf{k}_{j}&=\sum\limits_{\xi = 1}^{N} \left(u_{j \xi} l_{\xi}\right) \mathbf{u}_{\xi}, \label{equation_kj}
\end{align}
where $\mathbf{k}_{j}$ denotes the $j$th column vector of the matrix $\mathbf{K}$.
In our implementation, we have adopted a modified version of the original CUR algorithm~\cite{Mahoney697}
such that the columns of $\mathbf{K}$ that are strongly correlated with the $N_{low}$ eigenvectors $\mathbf{u}_{\xi}$ with the smallest eigenvalues $l_\xi$  are disregarded.
This is achieved by defining the leverage scoring for each column of $\mathbf{K}$
\begin{align}
\omega_{j}&=\frac{1}{N_{low}} \sum\limits_{\xi=1}^{N} \gamma_{\xi j}, \label{equation_ls1}\\
\gamma_{\xi j}&= \begin{cases} u_{j\xi}^{2}, & \mbox{if } l_{\xi}/l_{\rm max}< \epsilon  \\ 0, & \mbox{otherwise} \end{cases} \label{equation_ls2}
\end{align}
where $l_{\rm max}$ is the maximum eigenvalue of $\mathbf{K}$ and $\epsilon$ is a parameter used to define the degree of rank compression.
By defining Eq.~\eqref{equation_ls1}, the $N_{low}$ columns of $\mathbf{K}$ and the local reference configurations corresponding to those columns
with largest leverage scorings are disregarded. The remaining ones are deemed to be the most important ones.
Our final representative structures are then those structures that contribute to these most important local reference configurations.

\begin{figure*}
\begin{center}
\includegraphics[width=0.85\textwidth, clip]{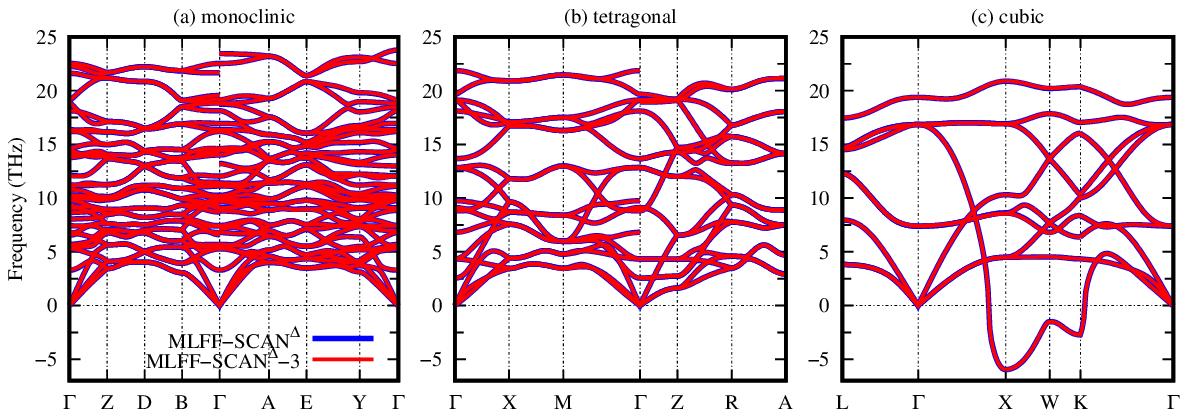}
\end{center}
\caption{Phonon dispersion relations of (a) monoclinic, (b) tetragonal, and (c) cubic  ZrO$_2$  at 0 K predicted by MLFF-SCAN$^\Delta$ (blue lines)
and MLFF-SCAN$^\Delta$-3 (red lines).
}
\label{fig:phonon_two_ML_SCAN_delta}
\end{figure*}

Using the SVD rank compression introduced above, we selected a subset of structures ($T^{(3)}$) from the $T^{(2)}$ dataset (including 1275 structures of 24 atoms).
We employed pair descriptors with low spatial resolution (0.8 $\AA$) and a small number of radial basis functions (8) to construct the kernel.
The actual number of SVD compressed structures depends on the parameter $\epsilon$ in Eq.~\eqref{equation_ls2}.
For instance, the resulting subset $T^{(3)}$ contains 168 structures, when $\epsilon=10^{-10}$.
This number will be reduced when $\epsilon$ is increased. As shown in Table~\ref{tab:error_MLFF-DFT_SM}, in addition to
the MLFF-SCAN$^\Delta$ that is discussed in the main text, we generated another two new MLFFs (denoted as MLFF-SCAN$^\Delta$-2 and MLFF-SCAN$^\Delta$-3)
via the $\Delta$-ML approach. These two MLFFs used just 102 and 72 structures for generating the MLFF-$\Delta$.
One can observe that even 72 structures are sufficient to machine-learn the differences and the resulting MLFF-SCAN$^\Delta$-3
is still accurate, showing only slightly larger validation errors in energies as compared to MLFF-SCAN$^\Delta$.
Moreover, we find that MLFF-SCAN$^\Delta$-3 predicts very similar lattice parameters, energy differences between the phases (not shown),
as well as the phonon dispersion relations (see Fig.~\ref{fig:phonon_two_ML_SCAN_delta}) as compared to MLFF-SCAN$^\Delta$.
The same observations also apply when generating MLFF-RPA$^\Delta$, so that in practice, similar results as in the main text could be obtained
with just half the number of RPA calculations.

\begin{table*}
\caption {The validation root-mean-square errors (RMSE) in energies per atom (meV/atom), forces (eV/\AA) and stress tensors (kbar)
for MLFF-SCAN, MLFF-SCAN$^\Delta$ and MLFF-SCAN$^\Delta$-$k$ ($k=$2 and 3).
The latter three MLFFs are obtained by the $\Delta$-ML approach and
differ in the level of SVD rank compression when constructing the $T^{(3)}$ dataset.
Here, the parameter $\epsilon$ defines the degree of rank compression [see Eq.~\eqref{equation_ls2}]  and $N_{\rm str}$
represents the number of SVD compressed structures used to machine-learn the differences.
The test dataset includes 120 structures of 96 atoms.
}
\begin{ruledtabular}
\begin{tabular}{lccccc}
 & Energy & Force & Stress & $\epsilon$ & $N_{\rm str}$ in $T^{(3)}$ \\
               \hline
MLFF-SCAN$^\Delta$ &  2.37 & 0.139 & 2.30 & 1E-10 & 168  \\
MLFF-SCAN$^\Delta$-2 &  2.45 & 0.139 & 2.28 & 1E-08 & 102  \\
MLFF-SCAN$^\Delta$-3 &  2.46 & 0.139 & 2.29 & 1E-07 &  72  \\
MLFF-SCAN            &  2.49 & 0.139 & 2.38 &       &      \\
\end{tabular}
\end{ruledtabular}
\label{tab:error_MLFF-DFT_SM}
\end{table*}

\newpage
\bibliographystyle{apsrev4-1}
\bibliography{reference} 